\documentclass[aip,jcp,11pt,reprint,floatfix]{revtex4-1}
\usepackage[utf8]{inputenc}
\usepackage{amsmath}
\usepackage{color}
\usepackage{graphicx}

\usepackage{algcompatible}
\usepackage{newfloat}

\DeclareFloatingEnvironment[
    fileext=loa,
    listname=List of Algorithms,
    name=ALGORITHM,
    placement=tbhp,
]{algorithm}

\newcommand{\ket}[1]{\left| #1 \right>} 
\newcommand{\bra}[1]{\left< #1 \right|} 
\newcommand{\matrixel}[3]{\left< #1 \vphantom{#2#3} \right|
 #2 \left| #3 \vphantom{#1#2} \right>} 
\newcommand{\bv}[1]{\ensuremath{\mathbf{#1}}} 

\begin{document}
\title{Finite-temperature coupled cluster: Efficient implementation and application to prototypical systems}
\date{March 2019}
\author{Alec F. White}
\email{whiteaf@berkeley.edu}
\author{Garnet Kin-Lic Chan}
\email{gkc1000@gmail.com}
\affiliation{Division of Chemistry and Chemical Engineering, California Institute of Technology, Pasadena,
California 91125, USA}

\begin{abstract}
We discuss the theory and implementation of the finite temperature coupled cluster singles and doubles (FT-CCSD) method including the equations necessary for an efficient implementation of response properties. Numerical aspects of the method including the truncation of the orbital space and integration of the amplitude equations are tested on some simple systems, and we provide some guidelines for applying the method in practice. The method is then applied to the 1D Hubbard model, the uniform electron gas at warm, dense conditions, and some simple materials. The performance on model systems at high temperatures is encouraging: for the 1-dimensional Hubbard model FT-CCSD provides a qualitatively accurate description of finite-temperature correlation effects even at $U = 8$, and it allows for the computation of systematically improvable exchange-correlation energies of the warm, dense UEG over a wide range of conditions. We highlight the obstacles that remain in using the method for realistic {\it ab initio} calculations on materials. 
\end{abstract}

\maketitle

\section{Introduction}
An {\it ab initio} description of the thermal properties of molecules and materials remains a significant challenge. In many cases, experimental temperatures are so small relative to the lowest energy electronic excitations that electronic temperature can be assumed to be effectively zero. However, there are cases where this assumption is not justified. Some examples include
\begin{enumerate}
    \item warm, dense matter
    \item the low-energy phases of correlated materials
    \item metallic systems
\end{enumerate}
In these systems, the electronic temperature cannot be ignored, and how best to incorporate thermal effects into computational methods for treating electron correlation is an open question.

In warm, dense matter, the thermal effects are comparable in magnitude to the effects of electron correlation.\cite{Fortov2009,Desjarlais2010} Conditions like this occur in planetary cores,\cite{Benuzzi-Mounaix2014} and can be realized in the laboratory with high intensity lasers.\cite{Ernstorfer2009,Fletcher2015} Finite temperature density functional theory (DFT) coupled with molecular dynamics (MD) for the nuclei is the most practical computational method for such systems.\cite{Wang2011,Hu2015,Zhang2017} However, the parameterization of finite temperature density functionals is a non-trivial problem,\cite{Perrot1984,Karasiev2014,Karasiev2016} and a variety of quantum Monte Carlo (QMC) methods have been developed with the goal of providing an accurate treatment of finite-temperature fermionic systems to aid in the development of finite-temperature functionals.\cite{Ceperley1991,Foulkes2001,Schoof2011,Blunt2014,Dornheim2015,Liu2018} In this context, the warm, dense uniform electron gas (UEG) has become an important system both as a benchmark for new methods and as an ingredient in the parameterization of finite temperature density functionals.\cite{Brown2013a,Brown2013,Sjostrom2013,Filinov2015,Schoof2015a,Schoof2015,Groth2016,Malone2016,Dornheim2016a,Dornheim2016,Dornheim2018}

The rich electronic phases of correlated materials also require a treatment of electron correlation at finite temperature.
Here, the low energy excitations typically involve the spin degrees of freedom and thus phase transitions can occur on the emergent
exchange coupling temperature (or lower) scales. Theoretical work has largely focused on model systems such as the Hubbard model.\cite{Hubbard1963,Haule2007,Yang2011,Gull2013,LeBlanc2015} For such lattice problems, a variety of methods including dynamical mean field theory (DMFT),\cite{Georges1996,Kotliar2006} the dynamical cluster approximation (DCA),\cite{Hettler1998,Jarrell2001} and finite temperature extensions to the density matrix renormalization group (DMRG)\cite{Wang1997,Verstraete2004,White2009} are commonly used.

In {\it ab initio} calculations on metals, DFT often offers a good description, and it is common practice to use a thermal smearing of the electron density to ease convergence of the Kohn-Sham equations.\cite{Doll1999,Aarons2016} Explicit treatment of electron correlation in metals beyond density functional theory is less common though GW theory has been applied to metals.\cite{Liu2016} {\it Ab initio} DMFT has been used to study correlated metallic systems, especially those which undergo a low-temperature phase transition due to electron correlation (see Section IV of Ref.~\onlinecite{Kotliar2006} for a review).

Problems like these have spurred a recent interest in extending  {\it ab initio} electronic structure methods to the case of finite electronic temperatures. The simplest methods in this hierarchy are thermal mean-field theories, Hartree-Fock (HF) theory\cite{Mermin1963} or DFT.\cite{Mermin1965} The goal is to develop hierarchies that mirror those at zero temperature and approach the thermal full configuration interaction (FCI)\cite{Kou2014} limit with polynomial scaling approaches. Examples include finite temperature extensions of perturbation theory,\cite{Hermes2015,Santra2017} configuration interaction (CI),\cite{Harsha2019} Green's function methods,\cite{Kananenka2016,Welden2016,Neuhauser2017} or coupled cluster (CC) theory.\cite{Mandal2003,Hermes2015,White2018,Hummel2018,Harsha2019a}

The coupled cluster method is the method of choice for high-accuracy, ground-state, quantum chemistry calculations,\cite{Paldus1972,Cizek1980,Bartlett1981,Purvis1982,Crawford2000,Bartlett2007,Shavitt2009} and we believe it to be a promising method for finite temperature calculations as well. The first polynomial-scaling finite temperature generalization of coupled cluster theory was the thermal cluster cumulant theory of Mukherjee and coworkers.\cite{Sanyal1992,Sanyal1993,Mandal2001,Mandal2002,Mandal2003} Recently, there has been renewed interest in finite-temperature coupled cluster methods. Hermes and Hirata suggested a coupled cluster doubles method based on their "renormalized" perturbation theory,\cite{Hermes2015} White and Chan presented a finite-temperature extension of CCSD (FT-CCSD),\cite{White2018} Hummel published a finite temperature linearized, direct coupled cluster doubles method for periodic solids, and Harsha {\it et al} derived a finite temperature coupled cluster theory based on the thermofield formalism.\cite{Harsha2019a} Coupled cluster methods for the dynamics of finite temperature systems driven out of equilibrium have also been the subject of several recent studies.\cite{Dzhioev2015,White2019,Shushkov2019} Despite all this development, many practical questions remain unanswered, and it is the goal of this work to address such questions.

Working within the FT-CCSD formalism presented in Ref.~\onlinecite{White2018}, we will clarify several aspects of the theory and present the equations necessary for an efficient implementation of FT-CCSD including a response treatment of properties. In Section~\ref{sec:bench} we will discuss the numerical and computational aspects of the method in the context of some simple benchmark calculations. In Section~\ref{sec:app} we apply the method to several finite temperature systems. The 1D Hubbard model allows us to compare to exact results for different values of the onsite repulsion, and we find that FT-CCSD performs well even for $U = 8$, a relatively large value of the onsite repulsion. We present FT-CCSD calculations of the UEG exchange-correlation energy at finite temperature with particular emphasis on the potential of FT-CC methods to provide consistent, systematically improvable results over a wide range of temperatures and densities. Finally, some simple {\it ab initio} calculations on periodic solids serve to demonstrate both the potential of the method and the difficulties we face in {\it ab initio} calculations at a finite electronic temperature.

\section{Theory}
Here, we review and expand on the theory presented in Ref.~\onlinecite{White2018}. The theory is, in a fundamental sense, identical to the thermal cluster cumulant (TCC) theory of Mukherjee and coworkers,\cite{Sanyal1992,Sanyal1993,Mandal2001,Mandal2002,Mandal2003} but our focus is on using the FT-CCSD theory presented in Ref.~\onlinecite{White2018} as a computational tool.

\subsection{The FT-CC equations: Integral and differential forms}
The FT-CC contribution to the grand potential is determined from an integration in imaginary time,
\begin{equation}\label{eqn:Ecc}
    \Omega_{CC} = \frac{1}{\beta}\int_0^{\beta}d\tau \text{E}[\bv{s}(\tau)],
\end{equation}
where $\beta$ is the inverse temperature, $\bv{s}$ is a vector of FT-CC amplitudes, and the kernel, E, is local in imaginary time and given in Equation~\ref{eqn:E}.

The FT-CC equations can be derived directly from diagrammatics as in Ref.~\onlinecite{White2018}, or from the thermally normal-ordered ansatz of TCC. The amplitude equations are non-linear and, in integral form, are given by:
\begin{equation}\label{eqn:smu}
    s_{\mu}(\tau) = -\int_0^{\tau}d\tau'e^{\Delta_{\mu}
    (\tau' - \tau)}\text{S}_{\mu}[\bv{s}(\tau')].
\end{equation}
The index, $\mu$, runs over the amplitudes which are typically truncated at some excitation level. Here, the S kernel is local in imaginary time and is given in Appendix~\ref{sec:cc_equations} for the case of finite-temperature coupled cluster singles and doubles (FT-CCSD). $\Delta_{\mu}$ is the difference of orbital energies associated with the $\mu$th excitation. In Ref.~\onlinecite{White2018}, we chose to define the amplitudes, $s$, such that the occupation numbers were associated with each line appearing ``above" the interaction diagrammatically. Here, we adopt a slightly different convention where the occupation numbers are split symmetrically. For example, at first order, the definition of $s_i^a(\tau)$ differs between Ref.~\onlinecite{White2018} and this work,
\begin{align}
    \text{Ref.~\onlinecite{White2018}:} \qquad &f_{ai}(1 - n_a) \\
    \text{This work: } \qquad &f_{ai}\sqrt{n_i(1 - n_a)}
\end{align}
where $n$ is  Fermi-Dirac occupation number and $f$ is the finite temperature Fock matrix. The appropriate modifications to the amplitude and energy equations are shown in Appendix~\ref{sec:cc_equations}. {\it This modification changes neither the theory nor the results}, but it allows us to form effective integrals which retain the symmetry of the underlying integrals, and it leads to a more symmetric treatment of the $\lambda$ amplitudes.

In order to efficiently compute properties, we define a variational Lagrangian,
\begin{widetext}
\begin{equation}\label{eqn:Lagrangian}
	\mathcal{L} \equiv \frac{1}{\beta}\int_0^{\beta}d\tau \mathrm{E}(\tau) +
	\frac{1}{\beta}\int_0^{\beta} d\tau \lambda^{\mu}(\tau)\left[
	s_{\mu}(\tau) + \int_0^{\tau}d\tau' e^{\Delta_{\mu}(\tau' - \tau)} \mathrm{S}_{\mu}(\tau')\right].
\end{equation}
\end{widetext}
Note that this definition differs by a minus sign from that given in Ref.~\onlinecite{White2018}. This sign convention does not change the results, but makes the $\lambda$ equations more closely resemble those of the ground-state theory. The $\lambda$ amplitudes are defined by the condition that $\mathcal{L}$ is stationary with respect to variations of the $s$ amplitudes which leads to a linear equation:
\begin{equation}\label{eqn:lmu}
    \lambda^{\mu}(\tau) = -\mathrm{L}[\mathbf{s}(\tau),\tilde{\mathbf{\lambda}}(\tau)]
\end{equation}
We define the quantity, $\tilde{\lambda}$, as
\begin{equation}\label{eqn:ltilde}
    \tilde{\lambda}^{\mu}(\tau) \equiv \int_{\tau}^{\beta}d\tau'e^{\Delta_{\mu}(\tau - \tau')}\lambda^{\mu}(\tau').
\end{equation}
Given these integral equations (Equation~\ref{eqn:smu} and Equation~\ref{eqn:lmu}), one can easily obtain differential equations for $s$ and $\tilde{\lambda}$ directly:
\begin{align}
    \frac{ds_{\mu}}{d\tau} &= -\left\{\Delta_{\mu}s_{\mu}(\tau) + \mathrm{S}_{\mu}[\bv{s}(\tau)]\right\} \label{eqn:smud}\\
    \frac{d\tilde{\lambda}_{\mu}}{d\tau} &= \left\{\Delta_{\mu}\tilde{\lambda}_{\mu}(\tau) + 
    \mathrm{L}_{\mu}[\bv{s}(\tau),\tilde{\bv{\lambda}}(\tau)] \right\}.\label{eqn:lmud}
\end{align}

These equations, in integral (Equations~\ref{eqn:smu}, \ref{eqn:lmu}, and~\ref{eqn:ltilde}) or differential (Equations~\ref{eqn:smud} and~\ref{eqn:lmud}) form, are described in more detail in Appendix~\ref{sec:cc_equations} for the specific case of FT-CCSD. Once the $\lambda$ amplitudes have been computed, properties can be evaluated by computing the partial derivatives of the Lagrangian. In Section~\ref{sec:cc_response} we will show how these derivatives, including the response of the reference orbital energies, can be computed by contracting the basis representation of an operator with response densities.

\subsection{Choice of reference}
Like in zero temperature coupled cluster theory, the choice of reference orbitals will have some effect on the energy and properties. Unlike the ground state theory, the choice of {\it orbital energies} will also have an effect. In other words, for a given choice of orbitals, the relative partitioning of the energy between between 0th and 1st order will matter at finite temperature even though it does not at zero temperature. This difference is most easily conceptualized within the TCC formulation which uses a thermally normal ordered ansatz. At zero temperature, partitioning the orbitals into an occupied and virtual space defines entirely the normal-ordering with respect to that reference:
\begin{equation}
    N[ABC\ldots]_{T=0} = ABC\ldots - \langle ABC\ldots \rangle_{T=0} .
\end{equation}
This is because the expectation value in the zero temperature reference is determined entirely by the choice of occupied space. However, at some finite temperature ($T = T_0$), the normal-ordering depends on the occupations explicitly:
\begin{equation}
    N[ABC\ldots]_{T=T_0} = ABC\ldots - \langle ABC\ldots \rangle_{T=T_0}.
\end{equation}
This is because the thermal average will depend on the occupation numbers of the states in question which, in turn, are functions of the non-interacting, single-particle energies.

This means that we must always be careful to specify the reference energies as well as orbitals used for a particular calculation since it will affect the final answer. There are many possible choices of reference orbitals and energies, and some aspects of the choice of reference have been described by Sanyal {\it et al}.\cite{Sanyal1994}

\subsection{Numerical integration and propagation in imaginary time}\label{sec:quadrature}
In practice the imaginary-time integral to to determine the free energy must be done by numerical quadrature:
\begin{equation}\label{eqn:g}
	\int_0^{\beta}I(\tau)d\tau \approx \sum_x g^xI(\tau_x).
\end{equation}
The values of amplitudes at some finite set of $n_g$ points are stored and the tensor $g$ contains the quadrature weights. If the integral form of the equations are used (see Equations~\ref{eqn:smu}, \ref{eqn:lmu}, and~\ref{eqn:ltilde}), then the amplitudes are determined by solving an integral equation of the form
\begin{equation}\label{eqn:G}
	s(\tau_y) \sim \int_0^{\tau_y}I(\tau)d\tau \approx \sum_x G^x_y I(\tau_x)
\end{equation}
where $G$ is a tensor of quadrature weights and the integrand $I$ depends on the amplitudes. On the other hand, if the differential form of the equations
are used (see Equations~\ref{eqn:smud} and~\ref{eqn:lmud}), then the amplitudes are propagated like
\begin{equation}
    s(\tau_y) = s(\tau_{y - 1}) + \Delta s,
\end{equation}
where the step, $\Delta s$, is determined either from a differential equation integrator such as a Runge-Kutta,\cite{Runge1895,Weisstein} Adams,\cite{Bashforth1883,Weissteina} or Crank-Nicolson\cite{Crank1947} method. There is a relationship between the integral and differential form of the equations in that any integral method defined by a set of quadrature rules encoded in $G$ should be equivalent to some, generally non-trivial, integrator. If $G$ has non-zero diagonal entries then the associated integral iteration is equivalent to an implicit propagation scheme, like Crank-Nicolson, and otherwise it will be equivalent to an explicit propagation scheme, like 4th order Runge-Kutta.

\subsection{Response properties}\label{sec:cc_response}
Properties in FT-CC theory are best computed from the response of the grand potential to a perturbation. This is most easily accomplished by computing analytic derivatives of the Lagrangian presented in Equation~\ref{eqn:Lagrangian}. The $\lambda$ amplitudes are computed such that this Lagrangian is stationary with respect to variations in the amplitudes, so we need not consider the response of the amplitudes directly, but there are still several types of response that must be considered. We will first consider the derivative with respect to a parameter $\alpha$, where $\alpha$ represents the coupling to some operator $X$. In this case are 3 types of terms:
\begin{enumerate}
    \item Terms resulting from the explicit dependence of the Hamiltonian on $\alpha$.
    \item Terms resulting from the dependence of the occupation numbers and orbital energies on $\alpha$
    \item Terms resulting from the dependence of the orbitals themselves on $\alpha$
\end{enumerate}
Unlike in ground state coupled cluster, the orbital energies and the 1-electron part of the perturbation appear separately in the Lagrangian. This means that properties will depend on the relative partitioning of $X$ into a part that is included in the orbital energies and a part that appears as part of the perturbation, e.g. for a one-electron $X$:
\begin{equation}
    X_{pq} = X^{(0)}_{q}\delta_{pq} + X^{(1)}_{pq}.
\end{equation} 
Terms of type 1 are the simplest and they may be efficiently computed by tracing $X^{(1)}$ with the unrelaxed, normal-ordered FT-CCSD 1-RDM, $\gamma_N$ (or the 2-RDM $\Gamma_N$ for 2-electron properties), as described in Appendix~\ref{sec:cc_rdensities}. We use the subscript $N$ to indicate that these densities represent the response only to the thermally normal ordered part of the operator. Terms of type 2 can be incorporated by tracing a diagonal matrix, $d$, with $X^{(0)}$. The computation of this quantity is also described in Appendix~\ref{sec:cc_rdensities}. The incorporation of the response of the orbital energies and occupation numbers is crucial to obtaining a density matrix that has a trace equal to the electron number computed as $-\partial \Omega/\partial \mu$. The orbital response (type 3) must be included to compute fully relaxed properties, and it is possible to also incorporate this into a fully relaxed density matrix. In this work, we ignore this contribution for several reasons. First, in most cases we use zero temperature orbitals which means that there will be no orbital response contribution to the energy, entropy, or number of electrons. Furthermore, for the UEG, this term is rigorously zero because the form of the orbitals is fixed by the translation invariance of the system. Finally, we suspect that, as in zero-temperature CCSD, the orbital contribution to most properties is small, though this should ultimately be verified numerically.
The entropy can be computed from the derivative with respect to $\beta$ for which there are additional terms that we must consider:
\begin{enumerate}\setcounter{enumi}{3}
    \item terms arising from the explicit dependence of the Lagrangian on $\beta$
    \item terms arising from the dependence of the quadrature weights on $\beta$
    \item terms arising from the positions of the grid points which depend on $\beta$
\end{enumerate}
Terms of type 4 are simply proportional to the value of the Lagrangian itself (Equation~\ref{eqn:diff4}).
Terms of type 5 are related ultimately to the dependence of the integration limits on $\beta$. Terms of type 6 can also be computed for a given discretization, though these terms will vanish in the limit of a dense grid (Equation~\ref{eqn:diff6}). Precise equations for all these terms are given in Appendix~\ref{sec:cc_rdensities}.

\section{Benchmarks}\label{sec:bench}
In this section we will use some simple benchmarks to suggest an answer to several practical questions. How severe an approximation is the truncation of the amplitudes based on small occupation numbers? What types of grids are most effective and how many grid points are necessary to obtain a desired accuracy? What computational resources are required to perform a given calculation? In exploring these questions, we will focus on two small systems: the beryllium atom in a minimal basis (STO-3G) at fixed $\mu$ ($\mu = 0$) , and the 14 electron, unpolarized UEG in a basis of 33 plane wave orbitals at a fixed average number of electrons i.e. $\mu$ is adjusted such that $\langle N\rangle=14$. 

\subsection{Restricted occupied and virtual spaces}
One of the simplest ways to reduce the cost of FT-CCSD is to allow nonzero amplitudes only when the ``occupied" (``virtual") indices are associated with orbitals that have particle (hole) occupation number greater than some threshold. In general such a truncation will lead to approximate results, and we must ask what error is incurred and what kind of thresholds are acceptable.

Though it is tempting to assume that the contribution to the free energy due to excitation from an orbital with occupation $n_i$ is proportional to $n_i$, this is unfortunately not the case, and we must be careful when truncating the excitation space in this way. As a rough estimate, consider the 2nd order contribution to the grand potential due a 1-particle matrix element $v_{ai}$:
\begin{equation}
    \Omega_{ai}^{(2)} = \frac{1}{\beta} n_i(1 - n_a)|v_{ai}|^2\left[\frac{\beta}{\varepsilon_i - \varepsilon_a} + \frac{1 - e^{\beta(\varepsilon_i - \varepsilon_a)}}{\varepsilon_i - \varepsilon_a}\right].
\end{equation}
For the purpose of this analysis, we will assume, without loss of generality,  that $\mu = 0$. Consider the case where $\beta$ and/or $\varepsilon_i$ are large such that
\begin{equation}
    n_i \sim e^{-\beta\varepsilon_i}.
\end{equation}
If we furthermore assume that $v_{ai}$ and $(1 - n_a)$ are of order 1 and $\varepsilon_a$ is small, then we can extract the asymptotic behavior of the 2nd order expression, and we find that 
\begin{equation}
    \Omega_{ai}^{(2)} \sim \frac{1}{\beta\varepsilon_i^2}.
\end{equation}

\begin{figure}[!ht]
    \centering
    \includegraphics[scale=1.0]{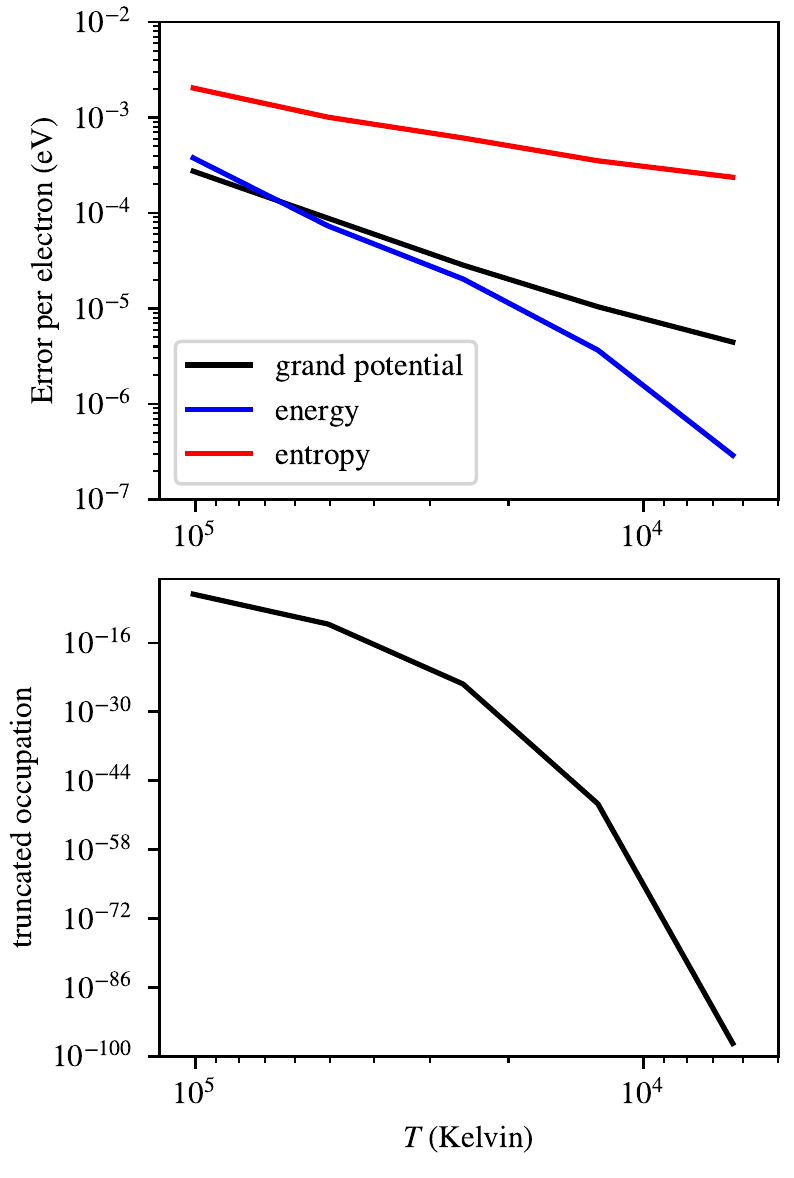}
    \caption{Error per electron due to truncating the ``virtual" space of the the minimal basis (STO-3G) Be atom so as not to include the lowest energy orbital. (Note that in the finite-temperature theory, the virtual space includes all orbitals, the virtual labels serving only to indicate
    the time-direction of the propagator associated with the orbital). The lower panel shows the hole occupation, $n_a - 1$, of this orbital. While the occupation decreases exponentially with temperature, the corresponding contribution to the properties decreases like some polynomial in temperature.}
    \label{fig:Be_athrsh}
\end{figure}

This would seem to suggest that, as a rigorous threshold, one should assume an error that goes like the natural log of the occupation numbers, $(-\varepsilon_i\ln n_i)^{-1}$, and not the occupation numbers themselves. This behavior is shown in Figure~\ref{fig:Be_athrsh} for the beryllium atom.
This figure clearly shows that the error incurred by truncating the cluster amplitudes is some polynomial in the inverse temperature even though the occupation numbers themselves decay exponentially. Despite this fact, we have observed that a threshold of approximately $1\times 10^{-30}$ is sufficient to guarantee errors of less than 1 meV per electron relative to the full FT-CCSD. In general this may be system dependent and it is always prudent to examine the convergence of relevant properties with respect to this threshold.

\subsection{Numerical integration and propagation schemes}
Efficient FT-CCSD calculations are critically dependent on the numerical quadrature used to compute the grand potential and the integral or differential scheme used to solve for the amplitudes. In this section we will discuss two questions:
\begin{enumerate}
    \item How does the error depend on the number of grid points for some simple numerical schemes?
    \item How many grid points are typically required?
\end{enumerate}
One has great freedom in choosing a propagation scheme and numerical integration scheme, and we cannot claim that the methods we use in this work are optimal. We believe that the best choice will ultimately be an adaptive scheme, but this is beyond the scope of this work.

In this work, we use a quadrature generated by Simpson's rule,\cite{Weissteinb} and use either the implicit integral method generated by the same Simpson's rule or an explicit Runge-Kutta (RK) propagator to compute the amplitudes. Even though we use the same grid for the quadrature that determines the grand potential and the integrator that determines the amplitudes, these are really separate sources of numerical error.
\begin{figure}[!ht]
    \centering
    \includegraphics{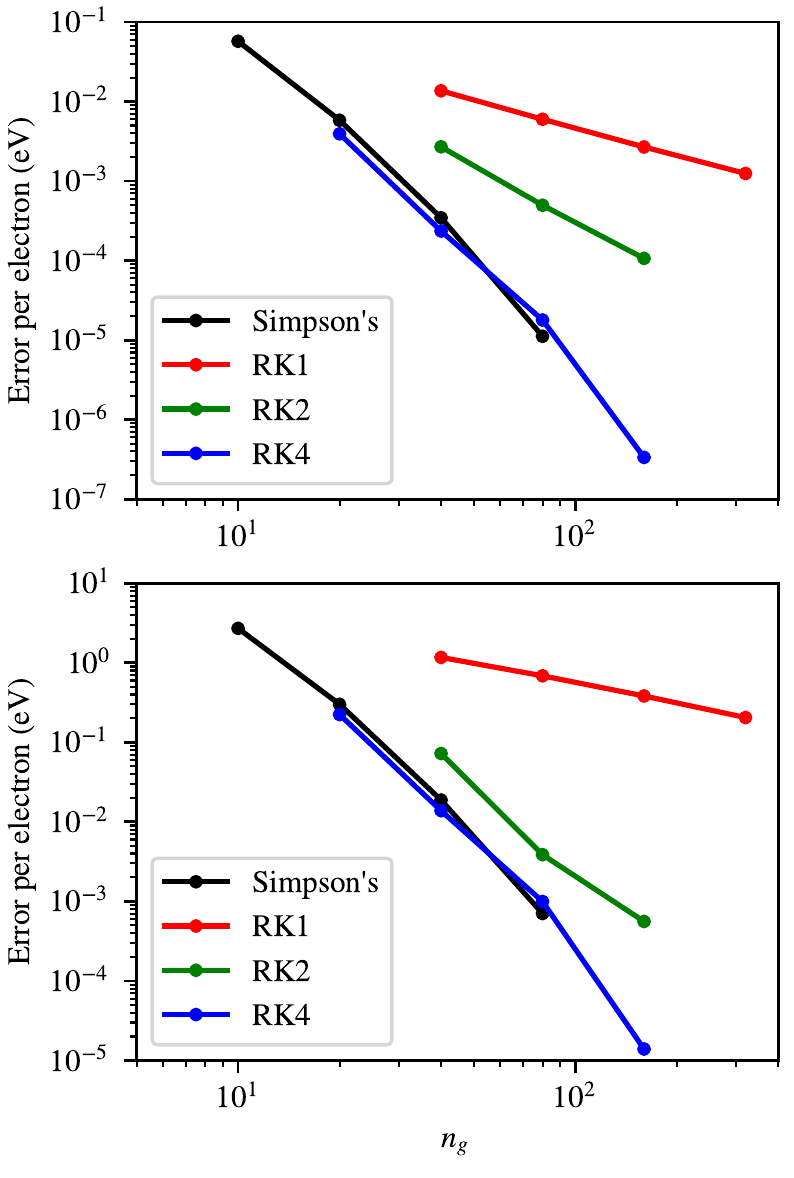}
    \caption{The error per electron in the exchange-correlation energy (top) and exchange-correlation entropy (bottom) for different integrators as a function of the number of grid points. The polynomial behavior of the error in the grid spacing matches the theoretical expectation (see Table~\ref{tab:numerr}).}
    \label{fig:ueg_grids}
\end{figure}
In Figure~\ref{fig:ueg_grids}, we show the error in the exchange-correlation energy and exchange-correlation entropy of the 14 electron unpolarized UEG in 33 plane-wave orbitals. The density is characterized by $r_s = 4$, and the reduced temperature is $\theta = 0.125$. Note that in all cases, the numerical error is controllable. The error in the entropy is larger than the error in the energy, and this is to be expected because
\begin{equation}
    S = -\beta (\Omega - E - \mu N).
\end{equation}
Therefore any error in the energy should be magnified in the entropy by a factor of the inverse temperature. This is consistent with the behavior observed in Figure~\ref{fig:Be_athrsh}. One must be cautious in using Figure~\ref{fig:ueg_grids} to suggest the ``best" numerical method. Though the Simpson's rule integrator may provide the smallest error for a given number of grid points, this implicit method requires the iterative solution of a non-linear equation at each step and is therefore considerably more expensive. We will return to this point in Section~\ref{sec:comp}.

The theoretical and observed behavior of the numerical error is summarized for different methods in Table~\ref{tab:numerr}.
\begin{table}[ht]
\begin{tabular}{c|c|c|c|c}
\hline\hline
     solver & Int. error & amp. error& obs. ($E_{xc}$)& obs. ($S_{xc}$)\\ \hline
     Simpson's & $h^4$ & $h^4$ &$h^{4.1}$  &$h^{4.0}$ \\
     RK1 & $h^4$ & $h^1$ &$h^{1.2}$  &$h^{0.8}$ \\
     RK2 & $h^4$ & $h^2$ &$h^{2.4}$  &$h^{2.8}$ \\
     RK4 & $h^4$ & $h^4$ &$h^{4.4}$  &$h^{4.6}$ \\
\hline\hline
\end{tabular}
\caption{\label{tab:numerr} Theoretical vs observed asymptotic error with respect to the step size, $h = 1/n_g$. The first column indicates the method used to compute the amplitudes, the second indicates the asymptotic behavior of the error due to the quadrature used to compute the grand potential and its derivatives, the third column shows the asymptotic behavior of the error due to the numerical solution of the amplitudes themselves. The final two columns show the behavior observed in Figure~\ref{fig:ueg_grids} for the error in the exchange-correlation energy and entropy respectively.}
\end{table}
This confirms that these numerical methods are behaving as expected and provides an answer to question (1). Furthermore, the clear asymptotic behavior of the numerical error allows us to estimate the numerical error in a given calculation and extrapolate the dense grid limit if desired. The deviations of the asymptotic error from its expected behavior in some cases may be due to the number of electrons not being sufficiently converged. For this system we fixed the number of electrons separately for each number of grid points to better than $1\times 10^{-4}$. In practice, the answer to question (2) can be obtained by monitoring the change in properties of interest as the number of grid points is increased. We have observed the number of grid points necessary for a given accuracy to scale roughly with $\beta$ with other parameters fixed.

\subsection{Timings and computational considerations}\label{sec:comp}
We must also consider the computational aspects of these calculations. For the numerical integration we can use either an explicit or implicit integrator in the solution of the amplitude equations. Implicit methods will generally be more accurate and more stable at the cost of iteratively solving a non-linear equation at each grid point. This trade-off is illustrated in Table~\ref{tab:timings} which suggests that explicit methods will usually be cheaper, though implicit methods may be preferable in some cases.
\begin{table}[ht]
    \centering
    \begin{tabular}{c|cc}
    \hline\hline
         Method & $n_g$ & time (s) \\
         \hline
         Simpson's & 40 & 4054\\
         RK1 &  320& 5720\\
         RK2 &  80& 2447\\
         RK4 &  40& 2128\\
    \hline\hline
    \end{tabular}
    \caption{Minimum number of grid points (given calculations for $n_g = 20,40,80,160,320$) necessary to obtain sub-millivolt error per electron in the exchange-correlation energy for the 14 electron UEG system, and the time of that calculation. All calculations were performed on a single 28 core node. }
    \label{tab:timings}
\end{table}
In particular, we have observed that at lower temperatures the differential equations can become ``stiff." In such cases, the step-size necessary to stably integrate the equations with an explicit method may be impractically small and an implicit integrator may be more efficient.

Though the differential and integral form of the amplitude equations do not differ conceptually, they suggest slightly different algorithms. The algorithm that mirrors the differential form of the algorithm is described in Algorithm~\ref{alg:diff}.
\begin{algorithm}\label{alg:diff}
\begin{algorithmic}[1]
\STATE{Initialize $s(\tau_0) = 0$}
\STATE{Initialize $\Omega_{CC} = 0$}
\FOR{$i = 1:n_g$}
    \STATE{Compute $\Delta s$ from Eqn.~\ref{eqn:smud}}
    \STATE{Form $s(\tau_i) = s(\tau_{i - 1}) + \Delta s$}
    \STATE{Increment $\Omega_{CC} =\Omega_{CC} + g_i\mathrm{E}[s(\tau_i)]/\beta$}
\ENDFOR
\end{algorithmic}
\caption{Solve for the amplitudes using the differential form of the equations. The key step is line 4 where either an explicit integrator (like RK4) or an implicit integrator (like the Crank-Nicolson method) is used to find the step.}
\end{algorithm}
The algorithm that follows the integral form of the equations is given in Algorithm~\ref{alg:int}.
\begin{algorithm}\label{alg:int}
\begin{algorithmic}[1]
\STATE{Initialize $s(\tau_0) = 0$}
\FOR{$i = 1:n_g$}
    \STATE{Compute S$[s(\tau_j)]$ for $j \leq i$}
    \STATE{Compute $s(\tau_i)$ from Eqn.~\ref{eqn:smu}}
\ENDFOR
\STATE{Compute $\Omega_{cc}$ from Eqn.~\ref{eqn:Ecc}}
\end{algorithmic}
\caption{Solve for the amplitudes using the integral form of the equations. The key step is line 4 where $s(\tau_i)$ is solved from Equation~\ref{eqn:smu} discretized as shown in Equations~\ref{eqn:G}. Depending on the form of the quadrature, this may or may not require the iterative solution of a system of non-linear equations.}
\end{algorithm}
In both cases, the most expensive step is is evaluation of the S kernel (Equations~\ref{eqn:ccS1} and~\ref{eqn:ccS2} for CCSD). The number of times that this kernel must be evaluated depends on the specific integrator or quadrature.

The computational scaling of FT-CCSD is asymptotically the same as for ground-state CCSD, but the prefactor is considerably larger due to the number of grid points and the fact that there is no distinction between ``occupied" and "virtual" orbital spaces. The additional memory cost due to the grid points can be ameliorated by using disk storage as shown in Table~\ref{tab:impl}.
\begin{table}[ht]
    \centering
    \begin{tabular}{c|ccc}
    \hline\hline
    method & disk & mem. & cpu \\ \hline
    incore & - & $n_gN^4$ & $n_gN^6$\\
    disk & $n_gN^4$ & $N^4$ & $n_gN^6$\\
    \hline\hline
    \end{tabular}
    \caption{Scaling of disk storage, memory, and cpu time for the fully-incore and disk-based implementations of FT-CCSD. $N$ indicates the number of orbitals and $n_g$ the number of grid points.}
    \label{tab:impl}
\end{table}
Technical improvements, such as distributed memory parallelization, are necessary to improve the performance further.

\section{Applications}\label{sec:app}
In order to demonstrate some features of the FT-CCSD method, we will now apply it to several prototypical systems.

\subsection{The Hubbard model}
First we consider the one-dimensional (1D) Hubbard model,\cite{Hubbard1963} an exactly solvable model of strong correlation. The one-band, 1D Hubbard model is given by the Hamiltonian
\begin{equation}
    H = -t\sum_{i\sigma} (a_{i,\sigma}^{\dagger}a_{i + 1,\sigma} + a_{i + 1,\sigma}^{\dagger}a_{i,\sigma})
     + U\sum_{i}n_{i\uparrow}n_{i\downarrow}
\end{equation}
where $i$ runs over the sites of a 1-dimensional lattice and $\sigma$ runs over the spin states of a spin-1/2 particle. The equilibrium properties of this model at finite temperature in the thermodynamic limit can be found exactly via the Bethe ansatz. This provides us with an opportunity to evaluate the strengths and weaknesses of FT-CCSD by comparing to an exact result for different values of the onsite repulsion $U/t$.

\begin{figure*}[ht]
    \centering
    \includegraphics{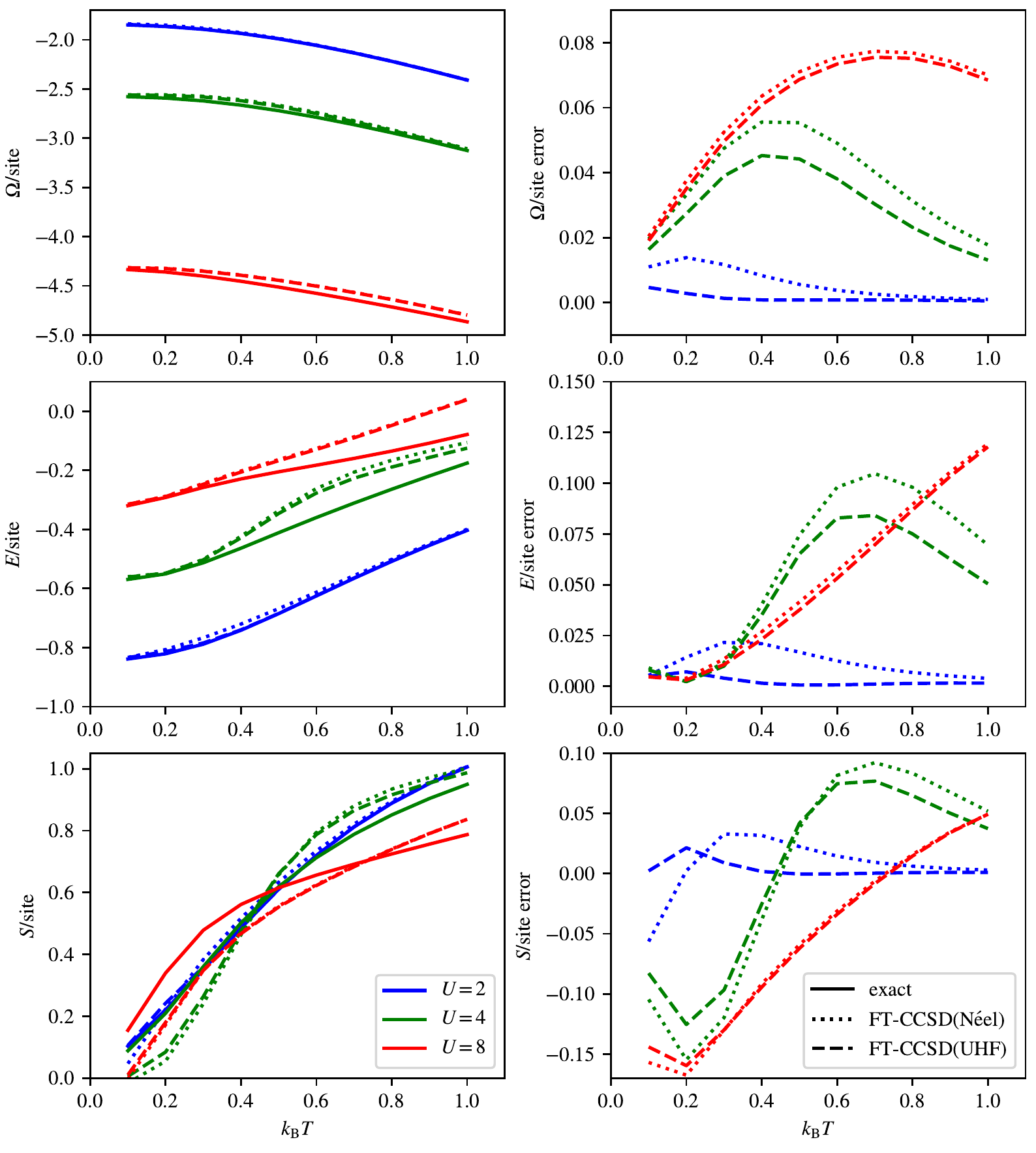}
    \caption{The grand potential per site (1st row), energy per site (2nd row), and entropy per site (3rd row) as a function of temperature for the 1D Hubbard model with $U = 2$ (blue), $U = 4$ (green), and $U = 8$ (red). In the first column we plot the exact results (solid line) and FT-CCSD results with a N\'{e}el state reference (dotted line) and a UHF reference (dashed line). In the second column the error in the FT-CCSD results is plotted. The entropy curves are also plotted in Appendix~\ref{sec:HubS} for clarity.}
    \label{fig:hubbard}
\end{figure*}
In Figure~\ref{fig:hubbard}  we show the exact and FT-CCSD grand potential, energy, and entropy per site for the 1D Hubbard model with periodic boundary conditions at half filling ($\mu = U/2$). The FT-CCSD results are taken from a 32 site lattice which is very close to the thermodynamic limit for the parameters considered here. FT-CCSD results from a N\'{e}el state reference and from a zero-temperature unrestricted Hartree-Fock (UHF) reference are shown to highlight the effect of different references. In both cases the reference non-interacting system is defined by the diagonal entries of the zero-temperature Fock matrix. The exact results are computed via the Bethe ansatz.\cite{Takahashi2002} In all cases, even for the relatively strongly correlated case of $U = 8$, FT-CCSD provides qualitatively correct results. The agreement with the exact result is better for smaller $U$ as we might expect from the performance of ground state CCSD on the Hubbard model.\cite{Paldus1984,Asai1999,LeBlanc2015} Though both sets of reference orbitals lead to FT-CCSD results with a similar level of accuracy, the optimized UHF orbitals clearly provide a better starting point. The difference between UHF and N\'{e}el orbitals gets smaller at larger $U$ where the UHF orbitals more closely resemble a N\'{e}el state.

\begin{figure}[!ht]
    \centering
    \includegraphics{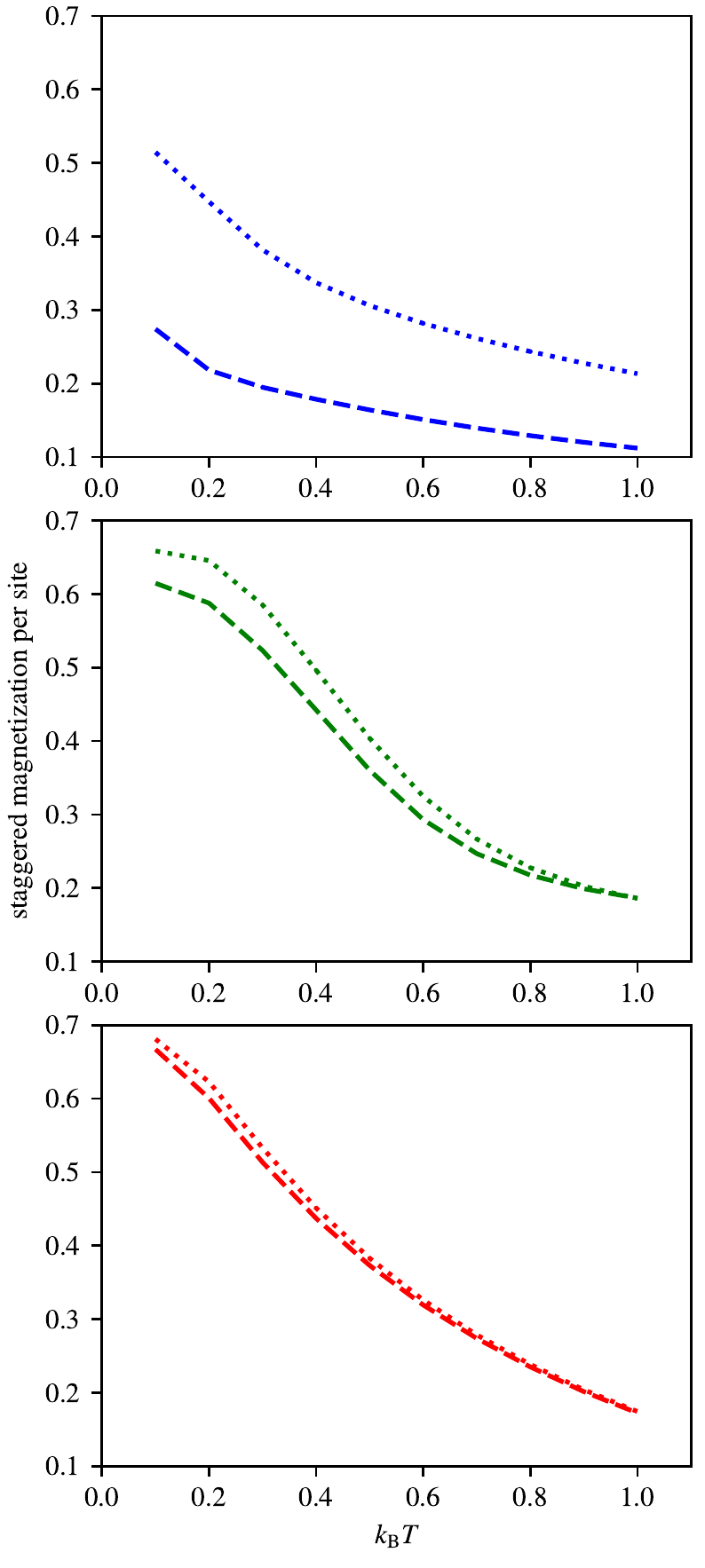}
    \caption{The staggered magnetization in the FT-CCSD 1-particle reduced density matrix as a function of temperature for 1D Hubbard model with $U = 2$ (top, blue), $U = 4$ (middle, green), and $U = 8$ (bottom, red). Results for both a N\'{e}el state reference (dotted line) and UHF reference (dashed line) are shown.}
    \label{fig:afm}
\end{figure}

The price that must be paid for coupled cluster calculations on strongly repulsive systems like this is artificial symmetry breaking in the reference orbitals. Both references (N\'{e}el and UHF) break spin symmetry, and this symmetry cannot be fully restored by FT-CCSD as shown in Figure~\ref{fig:afm} where we plot the staggered magnetization per site in the FT-CCSD 1-particle reduced density. This artificial symmetry breaking is not an issue if one is interested in just the energy or grand potential (see Figure~\ref{fig:hubbard}), but it will likely obscure certain types of phase transitions. The systematic underestimation of the entropy at low temperatures shown in row 3 of Figure~\ref{fig:hubbard} is also related to this artificial symmetry breaking since the configuration obtained by flipping all the spins of the reference is not well-described and cannot properly contribute to the entropy. 

It is encouraging that FT-CCSD provides qualitatively correct results even for relatively strongly correlated case of $U = 8$. However, the symmetry-broken references required to obtain these results suggest that FT-CCSD would not be appropriate for describing a phase transition, like the N\'{e}el transition in the 3-dimensional Hubbard model, that is governed by a {\it spontaneous} breaking of spin symmetry.

\subsection{The warm dense UEG}
The warm, dense UEG has been the focus of much work within the quantum Monte Carlo community with the focus being the accurate computation of the exchange-correlation energy.\cite{Sjostrom2013,Brown2013,Brown2013a,Schoof2015,Schoof2015a,Dornheim2015a,Dornheim2016,Malone2016,Dornheim2016a,Groth2016,Dornheim2018,Dornheim2019} The 66 electron unpolarized and 33 electron polarized UEG are the most commonly considered finite-size models. The work in this area is best summarized in Ref.~\onlinecite{Dornheim2018}. Our interest in the UEG is twofold: we compare to QMC results where accurate QMC results are available, and we evaluate the potential of FT-CC methods to provide results for some sets of parameters where reliable QMC calculations are more difficult. In particular note Figures 18-20 of Ref.~\onlinecite{Dornheim2018} where the state-of-the-art QMC calculations on these systems are summarized. We compare to the following finite temperature QMC methods: configuration path-integral Monte Carlo (CPIMC)\cite{Schoof2011}, density matrix quantum Monte Carlo (DMQMC)\cite{Blunt2014} with the initiator approximation (iDMQMC),\cite{Malone2016} permutation-blocking path integral Monte Carlo (PB-PIMC),\cite{Dornheim2015} and restricted path integral Monte Carlo (RPIMC).\cite{Ceperley1991} The warm-dense UEG can be completely characterized by its density (or Wigner-Seitz radius, $r_s$) and its temperature, $\theta$, given in units of the Fermi energy. In general, CPIMC and DMQMC are expected to perform better at high density (low $r_s$) while PB-PIMC and RPIMC are expected to be more reliable at low density (high $r_s$). All finite temperature QMC methods should be more reliable at higher temperatures for which the sign problem is less severe. We do not report results for $r_s < 0.5$ or for $\theta > 1$ since a variety of methods including FT-CCSD should be reliable in these limits.

In the FT-CCSD calculations shown here, there are two sources of error: the finite basis set and the neglect of high-order excitations (triples, quadruples, etc.). Additionally, FT-CC results in the grand canonical ensemble will differ from QMC calculations in the canonical ensemble for a finite number of electrons. In other words, the finite-size error will be different in canonical and grand canonical ensembles. In Appendix~\ref{sec:UEG_basis} we describe two methods of basis set extrapolation and comment on the magnitude of the basis set error in these calculations.
\begin{figure}[!ht]
    \centering
    \includegraphics{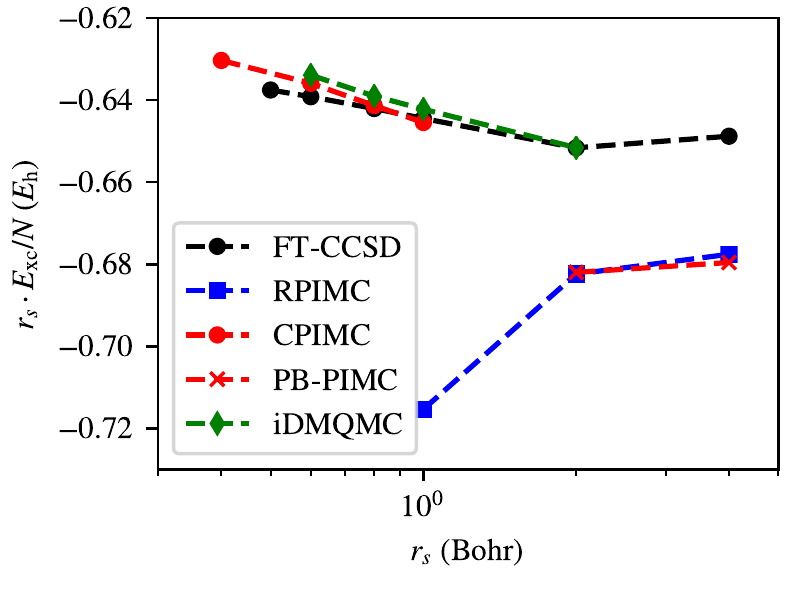}
    \caption{The exchange-correlation energy of the $N = 33$ polarized UEG as a function of $r_s$ for $\theta = 0.5$. The exchange-correlation energy is scaled by $r_s$ to make the scale of the plot more uniform. The FT-CCSD calculations are performed in a basis of 123 plane-waves. The close agreement between FT-CC and QMC approaches in this basis set is likely due to a favorable cancellation of errors.}
    \label{fig:PUEG_050}
\end{figure}
In Figures~\ref{fig:PUEG_050} and~\ref{fig:PUEG_025}, we show the exchange-correlation energy of the warm-dense polarized UEG as a function of $r_s$ as computed with FT-CCSD and a variety of QMC methods at reduced temperatures of 0.5 and 0.25 respectively. For $r_s \leq 2$ FT-CCSD agrees well with CPIMC and iDMQMC which should be reliable in this region (see Ref.~\onlinecite{Dornheim2018} section 5.7). For $r_s = 4$ FT-CCSD underestimates the magnitude of the exchange correlation at both temperatures shown here. This is likely due to the neglect of triples. At zero temperature, the triples are estimated to account for approximately 15$\%$ of the correlation energy at $r_s = 4$,\cite{Neufeld2017} and this is consistent with what we see in the warm dense regime.
\begin{figure}[!ht]
    \centering
    \includegraphics{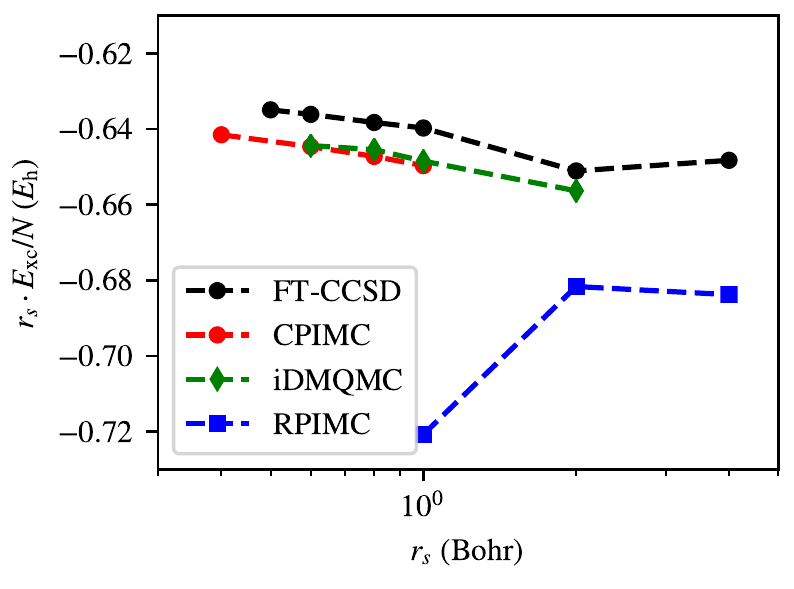}
    \caption{The exchange-correlation energy of the $N = 33$ polarized UEG as a function of $r_s$ for $\theta = 0.25$. The exchange-correlation energy is scaled by $r_s$ to make the scale of the plot more uniform. The FT-CCSD calculations are extrapolated to the complete basis set limit using the E1 method described in Appendix~\ref{sec:UEG_basis}. For $r_s = 4$ in particular the neglect of triple excitations is likeley the primary source of error.}
    \label{fig:PUEG_025}
\end{figure}
We expect the finite-basis error to be significant, especially at $\theta = 0.5$, and the good agreement at low $r_s$ for $\theta = 0.5$ is likely due to a cancellation of errors.

\begin{figure}[!ht]
    \centering
    \includegraphics{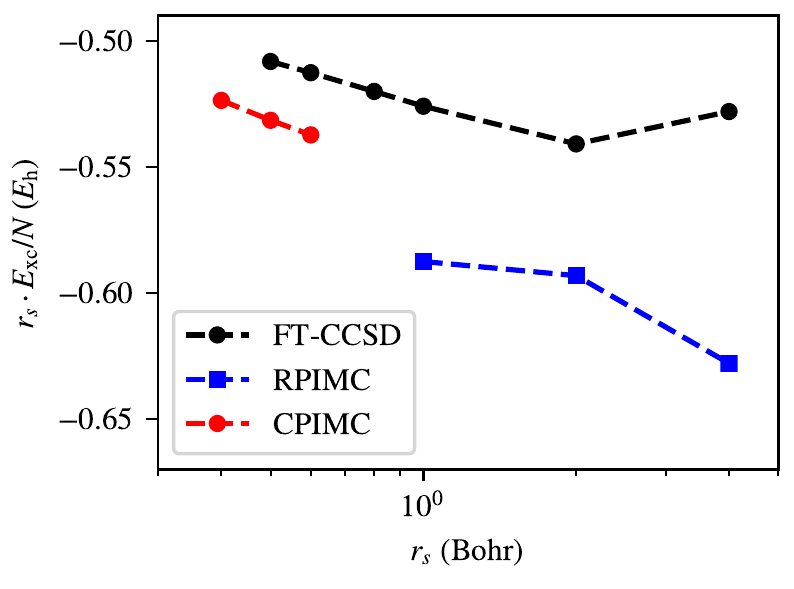}
    \caption{The exchange-correlation energy of the $N = 66$ unpolarized UEG as a function of $r_s$ for $\theta = 0.5$. The exchange-correlation energy is scaled by $r_s$ to make the scale of the plot more uniform. The FT-CCSD calculations are performed in a basis of 123 plane-waves.}
    \label{fig:UEG_050}
\end{figure}

\begin{figure}[!ht]
    \centering
    \includegraphics{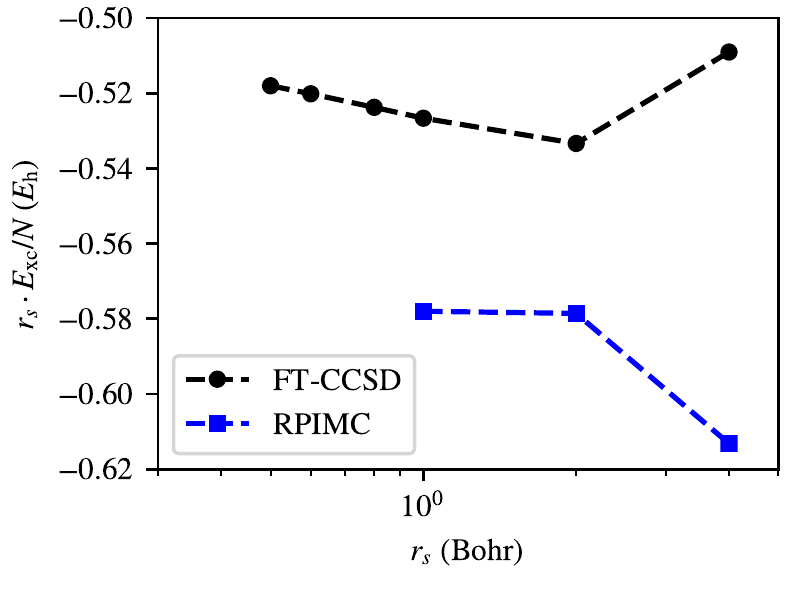}
    \caption{The exchange-correlation energy of the $N = 66$ unpolarized UEG as a function of $r_s$ for $\theta = 0.25$. The exchange-correlation energy is scaled by $r_s$ to make the scale of the plot more uniform. The FT-CCSD calculations are performed in a basis of 123 plane-waves.}
    \label{fig:UEG_025}
\end{figure}

In Figures~\ref{fig:UEG_050} and~\ref{fig:UEG_025} we show analogous calculations for the $N = 66$ unpolarized UEG.
For $N = 66$, differences between grand canonical and canonical ensembles should be smaller, and we expect the primary source of error to be the basis set for $r_s \leq 4$ and the neglect of higher excitations at $r_s = 4$. In Appendix~\ref{sec:UEG_basis} we provide an analysis of the finite basis error which supports this claim.

More detailed calculations are necessary to make definitive statements about this system. These include calculations in larger basis sets, calculations that allow for an estimate of triples, and calculations that provide an estimate of the finite-size error. FT-CC has the potential to provide systematically improvable results for the polarized and unpolarized UEG for $r_s \leq 4$ and for a very wide range of temperatures. For even moderate $r_s$ (such as $r_s<4$) it is known that zero-temperature mean-field theory gives a wide range of broken symmetry solutions~\cite{Loos2016} and the role of these broken symmetry states in subsequent coupled cluster calculations at finite temperature should be explored. Additionally, classifying correlation in terms of the order of the coupled cluster excitations can provide insight into the nature of correlation in this important system at finite temperature. 

\subsection{Ab initio Hamiltonians}
Finally, we consider the application of FT-CCSD to the {\it ab initio} problem. This problem is characterized by a number of difficulties including
\begin{itemize}
    \item converging to the thermodynamic limit in materials
    \item larger 1-particle basis sets and/or plane-wave cutoffs may be required at finite temperature because states with larger kinetic energy are populated
    \item the large number of grid points required to control the numerical error at lower temperatures
    \item the inclusion of finite temperature nuclear effects
\end{itemize}
FT-CCSD in its current form is still too expensive for us to meaningfully address all these difficulties, however we will nonetheless show that it is possible to apply FT-CCSD to the problem of {\it ab initio} calculations on materials within the framework of local basis functions. In the following calculations, we use a minimal valence basis set of periodic Gaussian orbitals (SZV)\cite{VandeVondele2007} and GTH pseudopotentials.\cite{Goedecker1996,Hartwigsen1998} The matrix elements have been obtained from the PySCF program package\cite{Sun2018} using plane-wave density fitting.\cite{VandeVondele2005} Zero-temperature, ground-state CCSD calculations were performed as described in Ref.~\onlinecite{Mcclain2017}.

\begin{figure}[!ht]
    \centering
    \includegraphics{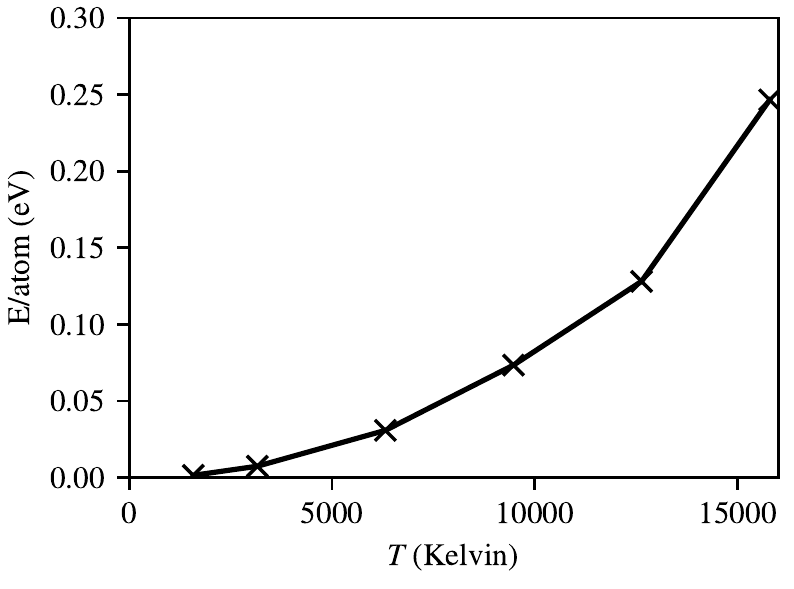}
    \caption{The FT-CCSD energy per atom of diamond relative to the zero temperature CCSD energy in the same basis.}
    \label{fig:diamond}
\end{figure}

\begin{figure}[!ht]
    \centering
    \includegraphics{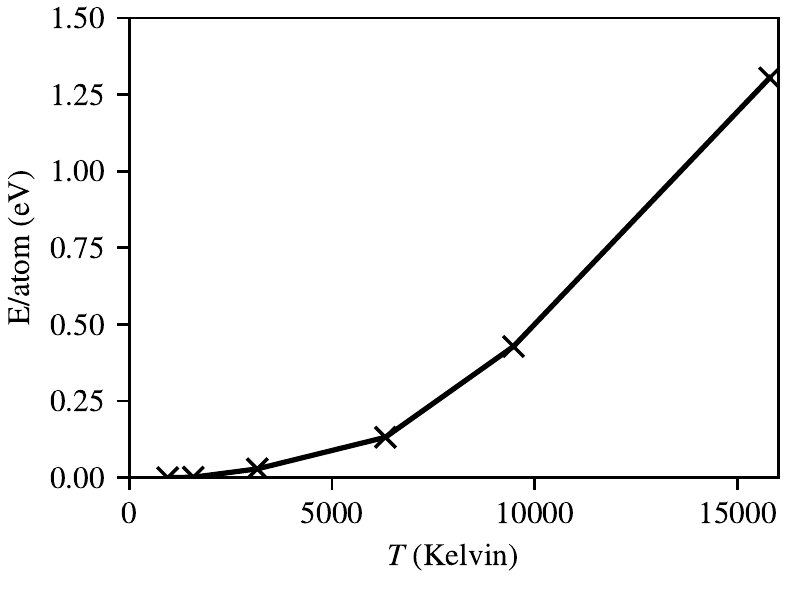}
    \caption{The FT-CCSD energy per atom of silicon relative to the zero temperature CCSD energy in the same basis.}
    \label{fig:silicon}
\end{figure}

In Figures~\ref{fig:diamond} and~\ref{fig:silicon} we show the energy per atom of diamond and silicon respectively relative to the zero temperature CCSD energy in the same basis set.
In both of these calculations twist averaging over a 3x3x3 k-point grid at fixed $\mu$ was used to partially alleviate finite-size errors, and a zero-temperature Hartree-Fock reference was used. As the temperature approaches zero, the FT-CCSD energy approaches the zero-temperature, ground-state CCSD energy. The difference between the ground-state and finite-temperature energy is more pronounced for silicon relative to diamond because silicon has lower energy excited states.

\begin{figure}[!ht]
    \centering
    \includegraphics{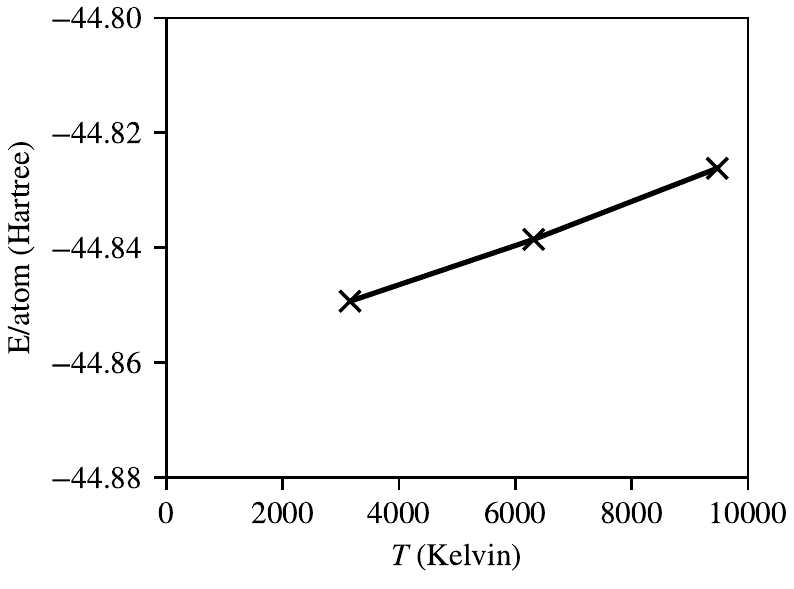}
    \caption{The FT-CCSD energy ($E_{\mathrm{h}}$) per atom of copper. A finite-temperature Hartree-Fock reference with $T = 0.01 E_{\mathrm{h}}$ was used for all points.}
    \label{fig:copper}
\end{figure}

Unfortunately, at lower temperatures, large orbital energy differences make integrating the FT-CCSD differential equations numerically unstable. This makes larger calculations difficult at lower temperatures. For example, for a 2-atom supercell of copper metal twist averaged over 3x3x3 mesh of k points, we were unable to reliably integrate the FT-CCSD equations much below 3000K (see Figure~\ref{fig:copper}).
Dealing with this difficulty is the subject of current investigations, but the initial results are nonetheless promising. It is rare to see finite-temperature calculations on materials where electron correlation is treated beyond the level of DFT, and FT-CCSD should be capable of providing valuable insight for such systems.

\section{Conclusions}
In this paper, we have discussed several aspects of FT-CCSD. All equations necessary for an efficient implementation have been presented, and some simple benchmarks have been provided to address the error incurred by restricting the orbital spaces and by numerical treatments of the imaginary time integration. Finally, we have shown results for the 1D Hubbard model, the warm, dense UEG, and some simple {\it ab initio} Hamiltonians. The 1D Hubbard model and the warm, dense UEG have allowed us to evaluate the strengths and weaknesses of FT-CCSD relative to exact or nearly exact results, and we find that, broadly speaking, FT-CCSD performs well for systems in which we might expect CCSD to perform well at zero temperature. For the warm, dense UEG, more calculations are needed to make truly definitive estimates of the exchange-correlation energy, but FT-CCSD performs well over a wide range of temperatures and densities. {\it Ab initio} Hamiltonians present some difficulties because of the large system sizes necessary to approach the thermodynamic limit in materials applications, and because of the relatively low temperatures necessary to obtain results of relevance to many phenomena of interest. The results for small models of silicon and diamond clearly show that the ground state CCSD energy is the zero temperature limit of FT-CCSD, and we are currently pursuing solutions to the numerical problems at low temperatures.

Future work on finite-temperature coupled cluster methods is proceeding in three directions:
\begin{enumerate}
    \item Technical improvements to address larger systems
    \item Theoretical improvements and approximations to more reliably treat lower temperatures
    \item Applications: more precise calculations on the UEG, benchmark {\it ab initio} calculations on materials in the warm-dense regime, {\it ab initio} calculations of metallic systems at ambient temperatures.
\end{enumerate}

\begin{acknowledgements}
  This work is supported by the US Department of Energy, Office of Science, via grant number SC0018140. The finite-temperature
  CC code relies on the PySCF software framework. The mean-field and periodic software infrastructure in PySCF has been 
  developed with support from the US National Science Foundation under award no. 1657286.
  GKC was also supported by the Simons Foundation, via the Many-Electron Collaboration, and via the Simons Investigator program. AFW would like to thank Matthew Foulkes for helpful discussions, and Chong Sun for help with the thermodynamic Bethe ansatz.
\end{acknowledgements}

\appendix 

\section{FT-CCSD energy, amplitude, $\lambda$ equations}\label{sec:cc_equations}
We will now state the FT-CCSD energy, amplitude, and $\lambda$ equations. We will use ``thermal" 1-electron and 2-electron integrals integrals:
\begin{align}
    f_{ij} &\equiv \sqrt{n_in_j}\left[\matrixel{i}{f}{j} - \delta_{ij}\varepsilon_i\right]\\
    f_{ia} &\equiv \sqrt{n_i(1 - n_a)}\matrixel{i}{f}{a} \\
    f_{ai} &\equiv \sqrt{n_i(1 - n_a)}\matrixel{a}{f}{i} \\
    f_{ab} &\equiv \sqrt{(1 - n_a)(1 - n_b)}\left[
    \matrixel{a}{f}{b} - \delta_{ab}\varepsilon_b\right]
\end{align}
\begin{align}
    \bra{ij}\ket{ab} &\equiv \sqrt{n_in_j(1 - n_a)(1 - n_b)}\nonumber \\
    & \qquad \times \left[
    \matrixel{ij}{V}{ab} - \matrixel{ij}{V}{ba}\right]\\
    \bra{ij}\ket{ka} &\equiv \sqrt{n_in_jn_k(1 - n_a)}\nonumber \\
    & \qquad \times \left[
    \matrixel{ij}{V}{ka} - \matrixel{ij}{V}{ak}\right]\\
    \text{etc.} \nonumber 
\end{align}
The operator $f$ is the Fock operator of the finite-temperature mean-field density, the orbital energies, $\varepsilon_p$, define the mean field system, and $V$ is the 2-particle Coulomb interaction.

The FT-CCSD grand potential is computed as
\begin{equation}
    \Omega_{CC} = \frac{1}{\beta}\sum_y g_y \text{E}(\tau_y)
\end{equation}
where $g$ is the tensor of weights for some numerical integration scheme (see Section~\ref{sec:quadrature}), and the kernel, E, is given by
\begin{equation}\label{eqn:E}
	\mathrm{E}(\tau) \equiv \sum_{ia}f_{ia}s_i^a(\tau) + 
    \frac{1}{4}\sum_{ijab}\langle ij||ab\rangle 
    [s_{ij}^{ab}(\tau) + 2s_i^a(\tau)s_j^b(\tau)].
\end{equation}

The FT-CCSD amplitude and $\lambda$ iterations can be written as:
\begin{align}
	s_i^a(\tau_y) &= -\tilde{\text{S}}_i^a(\tau_y) \\
	s_{ij}^{ab}(\tau_y) &= -\tilde{\text{S}}_{ij}^{ab}(\tau_y) \\
    \lambda^i_a(\tau_x) &= - \text{L}^i_a(\tau_x)\\
    \lambda^{ij}_{ab}(\tau_x) &= - \text{L}^{ij}_{ab}(\tau_x)
\end{align}
If the integral form of the equations are solved, we will use the following quadrature approximations to the integrated quantities:
\begin{align}
    \tilde{\text{S}}_{\mu}(\tau_y) &\equiv \sum_x G_x^y e^{\Delta_{\mu}
	(\tau_x - \tau_y)}\mathrm{S}_{\mu}(\tau_x)\\
    \tilde{\lambda}^{\mu}(\tau_x) &\equiv 
    \sum_y g_y\frac{G_x^y}{g_x}e^{\Delta_{\mu}(\tau_x - \tau_y)}
    \lambda^{\mu}(\tau_y) 
\end{align}
If instead the differential form of the equations are propagated in imaginary time, the $s$ and $\tilde{\lambda}$ amplitudes are computed directly from the S and L kernels. In either case, the utility of these definitions lies in the fact that the S and L kernels are local in time and are closely related to the ground state CCSD equations. For the singles, we find that
\begin{widetext}
\begin{align}\label{eqn:ccS1}
	\text{S}_i^a(\tau_x) &= f_{ai} + \sum_b f_{ab}s_i^b(\tau_x) - \sum_j f_{ji}s_j^a(\tau_x) + 
	\sum_{jb}\bra{ja}\ket{bi}s_j^b(\tau_x) + \sum_{jb}f_{jb}
	s_{ij}^{ab}(\tau_x)\nonumber \\
	&+ \frac{1}{2}\sum_{jbc}\bra{aj}\ket{bc}s_{ij}^{bc}(\tau_x)
	- \frac{1}{2}\sum_{jkb}\bra{jk}\ket{ib}s_{jk}^{ab}(\tau_x)
	-\sum_{jb} f_{jb}s_i^b(\tau_x)s_j^a(\tau_x) + \sum_{jbc}\bra{ja}\ket{bc}
	s_j^b(\tau_x)s_i^c(\tau_x)  \nonumber \\
	&- \sum_{jkb}\bra{jk}\ket{bi}s_j^b(\tau_x)s_k^a(\tau_x)
	- \frac{1}{2}\sum_{jkbc}\bra{jk}\ket{bc}s_i^b(\tau_x)s_{jk}^{ac}(\tau_x)
	- \frac{1}{2}\sum_{jkbc}\bra{jk}\ket{bc}s_j^a(\tau_x)s_{ik}^{bc}(\tau_x) 
	\nonumber \\
	&+ \sum_{jkbc}\bra{jk}\ket{bc}s_j^b(\tau_x)s_{ki}^{ca}(\tau_x)
	+ \sum_{jkcd}\bra{jk}\ket{bc}s_i^b(\tau_x)s_j^c(\tau_x)s_k^a(\tau_x)
\end{align}
And similarly, for the doubles
\begingroup
\allowdisplaybreaks
\begin{align}\label{eqn:ccS2}
	\text{S}_{ij}^{ab}(\tau_x) &= \bra{ab}\ket{ij} + P(ij)\sum_{c}\bra{ab}\ket{cj}s_i^c(\tau_x)
	- P(ab)\sum_{k} \bra{kb}\ket{ij}s_k^a(\tau_x) + P(ab)\sum_{c}f_{bc}s_{ij}^{ac}(\tau_x) \nonumber \\ &- 
	P(ij)\sum_{k}f_{kj} s_{ik}^{ab}(\tau_x)
	+ \frac{1}{2}\sum_{cd}\bra{ab}\ket{cd}s_{ij}^{cd}(\tau_x)
	+\frac{1}{2}\sum_{kl}\bra{kl}\ket{ij}s_{kl}^{ab}(\tau_x)\nonumber \\ 
	&+ P(ij)P(ab)\sum_{kc}\bra{kb}\ket{cj}s_{ik}^{ac}(\tau_x) 
	+\frac{1}{2}P(ij)\sum_{cd}\bra{ab}\ket{cd}s_i^c(\tau_x)s_j^d(\tau_x)\nonumber \\
	&+ \frac{1}{2}P(ab)\sum_{kl}\bra{kl}\ket{ij}s_k^a(\tau_x)s_l^b(\tau_x) 
	- P(ij)P(ab)\sum_{kc}\bra{ak}\ket{cj}s_i^c(\tau_x)s_k^b(\tau_x)\nonumber \\
	&-P(ij)\sum_{kc}f_{kc}s_i^c(\tau_x)s_{kj}^{ab}(\tau_x)
	- P(ab)\sum_{kc}f_{kc}s_k^a(\tau_x)s_{ij}^{cb}(\tau_x)\nonumber \\ & + 
	P(ab)\sum_{kcd}\bra{ka}\ket{cd}s_k^c(\tau_x)s_{ij}^{db}(\tau_x) - 
	P(ij)\sum_{klc}\bra{kl}\ket{ci}s_k^c(\tau_x)s_{lj}^{ab}(\tau_x)\nonumber \\ & 
	+ P(ij)P(ab)\sum_{kcd}\bra{ak}\ket{cd}s_i^c(\tau_x)s_{kj}^{db}(\tau_x) - 
	P(ij)P(ab)\sum_{klc}\bra{kl}\ket{ic}s_k^a(\tau_x)s_{lj}^{cb}(\tau_x)\nonumber \\ &+ 
	\frac{1}{2}P(ij)\sum_{klc} \bra{kl}\ket{cj}s_i^c(\tau_x)s_{kl}^{ab}(\tau_x)
	- \frac{1}{2}P(ab)\sum_{kcd}\bra{kb}\ket{cd}s_k^a(\tau_x)s_{ij}^{cd}(\tau_x)\nonumber \\ &
	+ \frac{1}{4}\sum_{klcd} \bra{kl}\ket{cd}
	s_{ij}^{cd}(\tau_x)s_{kl}^{ab}(\tau_x)+ \frac{1}{2}P(ij)P(ab)\sum_{klcd}\bra{kl}\ket{cd}
	s_{ik}^{ac}(\tau_x)s_{lj}^{db}(\tau_x)\nonumber \\ &
	- \frac{1}{2}P(ab)\sum_{klcd} \bra{kl}\ket{cd}s_{kl}^{ca}(\tau_x)s_{ij}^{db}(\tau_x)
	- \frac{1}{2}P(ij)\sum_{klcd}\bra{kl}\ket{cd}s_{ki}^{cd}(\tau_x)s_{lj}^{ab}(\tau_x)
	\nonumber \\ &-
	\frac{1}{2}P(ij)P(ab)\sum_{kcd}\bra{kb}\ket{cd}s_i^c(\tau_x)s_k^a(\tau_x)s_j^d(\tau_x)
	+ \frac{1}{2}P(ij)P(ab)\sum_{klc}\bra{kl}\ket{cj}s_i^c(\tau_x)s_k^a(\tau_x)s_l^b(\tau_x)\nonumber\\
	&+\frac{1}{4}P(ij)\sum_{klcd}\bra{kl}\ket{cd}s_i^c(\tau_x)s_j^d(\tau_x)s_{kl}^{ab}(\tau_x)
	+ \frac{1}{4}P(ab)\sum_{klcd}\bra{kl}\ket{cd}s_k^a(\tau_x)s_l^b(\tau_x)s_{ij}^{cd}(\tau_x) \nonumber\\ &-
	P(ij)P(ab)\sum_{klcd}\bra{kl}\ket{cd}s_i^c(\tau_x)s_k^a(\tau_x)s_{lj}^{db}(\tau_x)
	- P(ij)\sum_{klcd}\bra{kl}\ket{cd}s_k^c(\tau_x)s_i^d(\tau_x) s_{lj}^{ab}(\tau_x) \nonumber \\
	&- P(ab)\sum_{klcd}\bra{kl}\ket{cd}s_k^c(\tau_x)s_l^a(\tau_x)s_{ij}^{db}(\tau_x)
	+\frac{1}{4}P(ij)P(ab)\sum_{klcd}\bra{kl}\ket{cd}
	s_i^c(\tau_x)s_k^a(\tau_x)s_l^b(\tau_x)s_j^d(\tau_x).
\end{align}
\endgroup

The kernel L is also local in time and is equal to the CCSD $\lambda$ equations evaluated with $\tilde{\lambda}$:
\begingroup
\allowdisplaybreaks
\begin{align}\label{eqn:ccL1}
	\text{L}_{a}^{i}(\tau_x) &= f_{ia} + \sum_b \tilde{\lambda}^i_b(\tau_x)f_{ba} - \sum_j \tilde{\lambda}^j_a(\tau_x)f_{ij}
	+ \sum_{jb}\tilde{\lambda}^j_b(\tau_x)\bra{bi}\ket{ja} + \sum_{jb}\bra{ij}\ket{ab}s_j^b(\tau_x)
	\nonumber \\
	&- \sum_{jb}\tilde{\lambda}^j_a(\tau_x) f_{ib}s_j^b(\tau_x) - \sum_{jb}
	\tilde{\lambda}^i_b(\tau_x) f_{ja}s_j^b(\tau_x) + \sum_{jbc}\tilde{\lambda}^i_c(\tau_x)\bra{cj}\ket{ab}s_j^b(\tau_x)\nonumber \\ &
	 -\sum_{jkb}\tilde{\lambda}^k_a(\tau_x)\bra{ij}\ket{kb}s_j^b(\tau_x)
	 + \sum_{jbc}\tilde{\lambda}^j_c(\tau_x)\bra{ci}\ket{ba}s_j^b(\tau_x) - \sum_{jkb}\tilde{\lambda}^k_b(\tau_x)\bra{ji}
	\ket{ka}s_j^b(\tau_x) \nonumber \\
	&- \frac{1}{2}\sum_{jkbc} \tilde{\lambda}^j_a(\tau_x)\bra{ik}
	\ket{bc}s_{jk}^{bc}(\tau_x)
	- \frac{1}{2}\sum_{jkbc}\tilde{\lambda}^i_b(\tau_x)\bra{jk}\ket{ac}
	s_{jk}^{bc}(\tau_x) + \sum_{jkbc}\tilde{\lambda}^j_b(\tau_x)\bra{ki}\ket{ca}s_{jk}^{bc}(\tau_x) \nonumber \\
	&- \sum_{jkbc} \tilde{\lambda}^j_a(\tau_x)\bra{ik}
	\ket{bc}s_j^b(\tau_x)s_k^c(\tau_x)
	- \sum_{jkbc}\tilde{\lambda}^i_b(\tau_x)\bra{jk}\ket{ac}
	s_j^b(\tau_x)s_k^c(\tau_x) \nonumber \\ &- \sum_{jkbc}\tilde{\lambda}^k_b(\tau_x)\bra{ji}\ket{ca}s_j^b(\tau_x)s_k^c(\tau_x)
	+ \frac{1}{2}\sum_{jbc}\tilde{\lambda}_{cb}^{ij}(\tau_x)\bra{cb}
	\ket{aj} \nonumber \\
	&- \frac{1}{2}\sum_{jkb}\tilde{\lambda}_{ab}^{kj}(\tau_x)\bra{ib}\ket{kj}
	- \sum_{jkbc}\tilde{\lambda}^{jk}_{ac}(\tau_x)\bra{ic}\ket{bk}
	s_j^b(\tau_x) - \sum_{jkbc}\tilde{\lambda}^{ik}_{bc}(\tau_x)\bra{jc}\ket{ak}s_j^b(\tau_x) \nonumber \\
	&+ \frac{1}{2}\sum_{jbcd}\tilde{\lambda}^{ij}_{cd}(\tau_x)\bra{cd}\ket{ab}s_j^b(\tau_x)
	+ \frac{1}{2}\sum_{jklb}\tilde{\lambda}^{kl}_{ab}(\tau_x)\bra{ij}\ket{kl}s_j^b(\tau_x)
	-\frac{1}{2}\sum_{jkbc}\tilde{\lambda}^{jk}_{ba}(\tau_x)f_{ic}s_{jk}^{bc}(\tau_x)\nonumber \\
	&- \frac{1}{2}\sum_{jkbc}\tilde{\lambda}^{ji}_{bc}(\tau_x)f_{ka}s_{jk}^{bc}(\tau_x)
	+\frac{1}{2}\sum_{jkbcd}\tilde{\lambda}^{jk}_{bd}(\tau_x)\bra{di}\ket{ca}s_{jk}^{bc}(\tau_x)
	-\frac{1}{2}\sum_{jklbc}\tilde{\lambda}^{jl}_{bc}(\tau_x)\bra{ki}\ket{la}s_{jk}^{bc}(\tau_x)
	\nonumber \\
	&+ \sum_{jkbcd}\tilde{\lambda}^{ji}_{bd}(\tau_x)\bra{kd}\ket{ca}s_{jk}^{bc}(\tau_x)
	- \sum_{jklbc}\tilde{\lambda}^{jl}_{ba}(\tau_x)\bra{ki}\ket{cl}s_{jk}^{bc}(\tau_x)
	- \frac{1}{4}\sum_{jkbcd}\tilde{\lambda}^{jk}_{ad}(\tau_x)\bra{id}\ket{bc}s_{jk}^{bc}(\tau_x)
	\nonumber \\ 
	&+\frac{1}{4}\sum_{jklbc}\tilde{\lambda}^{il}_{bc}(\tau_x)\bra{jk}\ket{al}s_{jk}^{bc}(\tau_x)
	-\sum_{jkbcd}\tilde{\lambda}^{ik}_{db}(\tau_x)\bra{dj}\ket{ac}s_j^b(\tau_x)s_k^c(\tau_x) \nonumber \\ &
	+ \sum_{jklbc}\tilde{\lambda}^{lk}_{ab}(\tau_x)\bra{ij}\ket{lc}s_j^b(\tau_x)s_k^c(\tau_x)
	- \frac{1}{2}\sum_{jkbcd}\tilde{\lambda}^{jk}_{ad}(\tau_x)\bra{id}\ket{bc}s_j^b(\tau_x)s_k^c(\tau_x) \nonumber \\ &
	+ \frac{1}{2}\sum_{jklbc}\tilde{\lambda}^{il}_{bc}(\tau_x)\bra{jk}\ket{ad}s_j^b(\tau_x)s_k^c(\tau_x)
	-\frac{1}{2}\sum_{jklbcd}
	\tilde{\lambda}^{kl}_{ca}(\tau_x)\bra{ij}\ket{db}s_j^b(\tau_x)s_{kl}^{cd}(\tau_x) \nonumber \\
	&- \frac{1}{2}\sum_{jklbcd}\tilde{\lambda}^{ki}_{cd}(\tau_x)\bra{lj}\ket{ab}s_j^b(\tau_x)s_{kl}^{cd}(\tau_x)
	- \sum_{jklbcd}\tilde{\lambda}^{jl}_{ad}(\tau_x)\bra{ik}\ket{bc}s_j^b(\tau_x)s_{kl}^{cd}(\tau_x)\nonumber \\ &
	-\sum_{jklbcd}\tilde{\lambda}^{il}_{bd}(\tau_x)\bra{jk}\ket{ac}s_j^b(\tau_x)s_{kl}^{cd}(\tau_x)
	+\frac{1}{4}\sum_{jklbcd}\tilde{\lambda}^{kl}_{ab}(\tau_x)\bra{ij}\ket{cd}s_j^b(\tau_x)s_{kl}^{cd}(\tau_x)\nonumber \\ &
	+\frac{1}{4}\sum_{jklbcd}\tilde{\lambda}^{ij}_{cd}(\tau_x)\bra{kl}\ket{ab}s_j^b(\tau_x)s_{kl}^{cd}(\tau_x)
	-\frac{1}{2}\sum_{jklbcd}\tilde{\lambda}^{kl}_{cb}(\tau_x)\bra{ji}\ket{da}s_j^b(\tau_x)s_{kl}^{cd}(\tau_x)
	\nonumber \\
	&-\frac{1}{2}\sum_{jklbcd}\tilde{\lambda}^{kj}_{cd}(\tau_x)\bra{li}\ket{ba}s_j^b(\tau_x)s_{kl}^{cd}(\tau_x)
	+ \frac{1}{2}\sum_{jklbcd}\tilde{\lambda}^{jl}_{ac}(\tau_x)
	\bra{ik}\ket{bd}s_j^b(\tau_x)s_k^c(\tau_x)s_l^d(\tau_x)\nonumber \\
	&+ \frac{1}{2}\sum_{jklbcd}\tilde{\lambda}^{il}_{bc}(\tau_x)
	\bra{jk}\ket{ad}s_j^b(\tau_x)s_k^c(\tau_x)s_l^d(\tau_x)
\end{align}
\endgroup

\begin{align}\label{eqn:ccL2}
	\text{L}_{ab}^{ij}(\tau_x) &= \bra{ij}\ket{ab} + P(ij)P(ab)f_{ia}\tilde{\lambda}^j_b(\tau_x) + 
	P(ij)\sum_c \tilde{\lambda}^i_c(\tau_x)\bra{cj}\ket{ab} \nonumber \\ &- P(ab)\sum_k\tilde{\lambda}^k_a(\tau_x)\bra{ij}
	\ket{kb}
	+ P(ij)P(ab)\sum_{kc}\tilde{\lambda}^j_b(\tau_x)\bra{ik}\ket{ac}
	s_k^c(\tau_x) \nonumber \\ &- P(ij)\sum_{kc}\tilde{\lambda}^i_c(\tau_x)\bra{kl}\ket{ab}s_k^c(\tau_x) - P(ab)\sum_{kc}
	\tilde{\lambda}^k_a(\tau_x)\bra{ij}\ket{cb}s_k^c(\tau_x) \nonumber \\
	&+ P(ab)\sum_c \tilde{\lambda}^{ij}_{ac}(\tau_x)f_{cb} 
	- P(ij)\sum_k\tilde{\lambda}^{ik}_{ab}(\tau_x)f_{jk} 
	+ \frac{1}{2}\sum_{cd} \tilde{\lambda}^{ij}_{cd}(\tau_x) \bra{cd}\ket{ab} \nonumber \\ &+
	\frac{1}{2}\sum_{kl}\tilde{\lambda}^{kl}_{ab}(\tau_x)\bra{ij}\ket{kl}
	+ P(ij)P(ab)\sum_{kc}\tilde{\lambda}^{ik}_{ac}(\tau_x)\bra{cj}\ket{kb}\nonumber \\ &
	-P(ij)\sum_{kc}\tilde{\lambda}^{ik}_{ab}(\tau_x)f_{jc}s_k^c(\tau_x)
	- P(ab)\sum_{kc}\tilde{\lambda}^{ij}_{ac}(\tau_x)f_{kb}s_k^c(\tau_x)\nonumber \\
	&+ P(ab)\sum_{kcd}\tilde{\lambda}^{ij}_{ad}(\tau_x)\bra{dk}\ket{bc}s_k^c(\tau_x) - 
	P(ij)\sum_{klc}\tilde{\lambda}^{il}_{ab}(\tau_x)\bra{jk}\ket{lc}s_k^c(\tau_x) \nonumber \\ &
	+ P(ij)P(ab)\sum_{kcd}\tilde{\lambda}^{ik}_{ad}(\tau_x)\bra{dj}\ket{cb}s_k^c(\tau_x)
	- P(ij)P(ab)\sum_{klc}\tilde{\lambda}^{il}_{ac}(\tau_x)\bra{kj}\ket{lb}s_k^c(\tau_x) \nonumber \\ 
	&- \sum_{kcd}\tilde{\lambda}^{ij}_{cd}(\tau_x)\bra{kd}\ket{ab}s_k^c(\tau_x)
	+ \sum_{klc}\tilde{\lambda}^{kl}_{ab}(\tau_x)\bra{ij}\ket{cd}s_k^c(\tau_x) \nonumber \\
	&- P(ij)\frac{1}{2}\sum_{klcd}\tilde{\lambda}^{ik}_{ab}(\tau_x)
	\bra{jl}\ket{cd}s_{kl}^{cd}(\tau_x) - P(ab)\frac{1}{2}\sum_{klcd}\tilde{\lambda}^{ij}_{ac}(\tau_x)
	\bra{kl}\ket{bd}s_{kl}^{cd}(\tau_x)\nonumber \\ &+ P(ij)P(ab)\sum_{klcd}\tilde{\lambda}^{ik}_{ac}(\tau_x)
	\bra{lj}\ket{db}s_{kl}^{cd}(\tau_x)
	- P(ab)\frac{1}{2}\sum_{klcd}\tilde{\lambda}^{kl}_{ca}(\tau_x)\bra{ij}\ket{db}s_{kl}^{cd}(\tau_x)\nonumber \\
	&- P(ij)\frac{1}{2}\sum_{klcd}\tilde{\lambda}^{ki}_{cd}(\tau_x)\bra{lj}\ket{ab}s_{kl}^{cd}(\tau_x)
	+ \frac{1}{4}\sum_{klcd}\tilde{\lambda}^{kl}_{ab}(\tau_x)\bra{ij}\ket{cd}s_{kl}^{cd}(\tau_x)
	\nonumber \\
	&+ \frac{1}{4}\sum_{klcd}\tilde{\lambda}^{ij}_{cd}(\tau_x)\bra{kl}\ket{ab}s_{kl}^{cd}(\tau_x)
	-P(ij)\sum_{klcd}\tilde{\lambda}^{ik}_{ab}(\tau_x)
	\bra{jl}\ket{cd}s_k^c(\tau_x)s_l^d(\tau_x) \nonumber \\ &- P(ab)\sum_{klcd}\tilde{\lambda}^{ij}_{ac}(\tau_x)\bra{kl}
	\ket{bd}s_k^c(\tau_x)s_l^d(\tau_x)
	- \sum_{klcd}\tilde{\lambda}^{ik}_{ad}(\tau_x)\bra{lj}\ket{cb}s_k^c(\tau_x)s_l^d(\tau_x) \nonumber \\
	&+ \frac{1}{2}\sum_{klcd}\tilde{\lambda}^{kl}_{ab}(\tau_x)\bra{ij}\ket{cd}s_k^c(\tau_x)s_l^d(\tau_x)
	+ \frac{1}{2}\sum_{klcd}\tilde{\lambda}^{ij}_{cd}(\tau_x)\bra{kl}\ket{ab}s_k^c(\tau_x)s_l^d(\tau_x)
\end{align}

\end{widetext}
In practice we use the intermediate scheme of Stanton and Gauss\cite{Stanton1991,Gauss1995} to compute S and L efficiently.

\section{FT-CCSD response densities}\label{sec:cc_rdensities}
As we discussed in Section~\ref{sec:cc_response}, the computation of derivatives can be efficiently implemented by computing response densities that can be contracted with the basis representation of operators to compute properties. In Section~\ref{sec:cc_response} we described a total of 6 terms relevant to the computation of derivatives. We will now describe how the contribution of all terms can be efficiently computed for FT-CCSD.

Terms of type 1 can be computed by evaluating the Lagrangian with the the quantities
\begin{equation}
    \frac{\partial E}{\partial \alpha}, \quad \frac{\partial \text{S}_{\mu}}{\partial \alpha},
\end{equation}
but it is more efficient to first form unrelaxed, normal-ordered 1- or 2-particle response densities, $\gamma_N$ or $\Gamma_N$, and then trace them with the appropriate operators when more than one property is desired. The expressions for these quantities are given in Equations~\ref{eqn:g1ia}-\ref{eqn:g2abij} with implied summations.
\begin{widetext}
The unrelaxed 1-RDM:
\begin{align}
	\frac{(\gamma_N)_{ia}}{\sqrt{n_i(1 - n_a)}}&=
	-g_y\tilde{\lambda}^i_a(\tau_y) \label{eqn:g1ia}\\
	\frac{(\gamma_N)_{ba}}{\sqrt{(1 - n_b)(1 - n_a)}} &= -g_y\tilde{\lambda}^i_a(\tau_y)s_i^b(\tau_y) - \frac{1}{2}g_y
	\tilde{\lambda}^{ki}_{cb}(\tau_y)s^{ca}_{ki}(\tau_y) \\
	\frac{(\gamma_N)_{ji}}{\sqrt{n_in_j}} &= g_y\tilde{\lambda}_a^j(\tau_y)s_i^a(\tau_y) + \frac{1}{2}g_y
	\tilde{\lambda}^{kj}_{ca}(\tau_y)s_{ki}^{ca}(\tau_y)\\
	\frac{(\gamma_N)_{ai}}{\sqrt{n_i(1 - n_a)}} &= g_ys_i^a(\tau_y) - g_y\tilde{\lambda}_b^j(\tau_y)s_{ji}^{ba}(\tau_y) +g_y\tilde{\lambda}^j_b(\tau_y)s_i^b(\tau_y)s_j^a(\tau_y)\nonumber \\
	&+ \frac{1}{2}g_y\tilde{\lambda}^{jk}_{bc}(\tau_y)
	s_i^b(\tau_y)s_{jk}^{ac}(\tau_y)
	+ \frac{1}{2}g_y\tilde{\lambda}^{jk}_{bc}(\tau_y)
	s_j^a(\tau_y)s_{ik}^{bc}(\tau_y)
\end{align}
The unrelaxed 2-RDM:
\begingroup
\allowdisplaybreaks
\begin{align}
	\frac{(\Gamma_N)_{ijab}}{\sqrt{n_jn_j(1 - n_a)(1 - n_b)}} &=
	    -g_y\tilde{\lambda}^{ij}_{ab}(\tau_y)\\
	\frac{(\Gamma_N)_{ciab}}{\sqrt{(1 - n_c)n_i(1 - n_a)(1 - n_b)}} &= 
	    -g_y\tilde{\lambda}^{ji}_{ab}(\tau_y)s_j^c(\tau_y)\\
	\frac{(\Gamma_N)_{jkai}}{\sqrt{n_jn_k(1 - n_a)n_i}} &= 
	    g_y\tilde{\lambda}^{jk}_{ab}(\tau_y)s_i^b(\tau_y)\\
	\frac{(\Gamma_N)_{cdab}}{\sqrt{(1 - n_c)(1 - n_d)(1 - n_a)(1 - n_b)}} &= 
	    -\frac{1}{2}g_y\tilde{\lambda}^{kl}_{ab}(\tau_y)s_{kl}^{cd}(\tau_y) 
	    - P(cd)g_y\frac{1}{2}\tilde{\lambda}^{kl}_{cd}(\tau_y)s_k^c(\tau_y)s_l^d(\tau_y)\\
	\frac{(\Gamma_N)_{bjia}}{\sqrt{(1 - n_b)n_jn_i(1 - n_a)}} &=     
	    -g_y\tilde{\lambda}^j_a(\tau_y)s_i^b(\tau_y) 
	    -g_y\tilde{\lambda}^{kj}_{ca}(\tau_y)s^{cb}_{ki}(\tau_y) 
	    +g_y\tilde{\lambda}^{kj}_{ac}(\tau_y)s_k^b(\tau_y)s_i^c(\tau_y)\\
	\frac{(\Gamma_N)_{klij}}{\sqrt{n_kn_ln_in_j}} &=
	    -\frac{1}{2}g_y\tilde{\lambda}^{kl}_{cd}(\tau_y)s^{cd}_{ij}(\tau_y) 
	    -P(ij)\frac{1}{2}g_y\tilde{\lambda}^{kl}_{cd}(\tau_y)s^c_i(\tau_y)s_j^d(\tau_y)\\
	\frac{(\Gamma_N)_{bcai}}{\sqrt{(1 - n_b)(1 - n)(1 - n_a)n_i}} &= 
	    -g_y\tilde{\lambda}^j_a(\tau_y)s_{ji}^{bc}(\tau_y) 
		-P(bc)g_y\tilde{\lambda}^j_a(\tau_y)s_j^b(\tau_y)s_i^c(\tau_y) \nonumber \\
	&-P(bc)\frac{1}{2}g_y\tilde{\lambda}^{lk}_{da}(\tau_y)s^{db}_{lk}(\tau_y)s_i^c(\tau_y)
		- P(bc)g_y\tilde{\lambda}^{kl}_{ad}(\tau_y)s_{il}^{cd}(\tau_y)s_k^b(\tau_y)\nonumber \\ 
	&+ \frac{1}{2}g_y\tilde{\lambda}^{kl}_{ad}(\tau_y)s_{kl}^{bc}(\tau_y)s_i^d(\tau_y) 
	    + g_y\tilde{\lambda}_{ad}^{kl}(\tau_y)s_k^b(\tau_y)s_i^d(\tau_y)s_l^c(\tau_y)\\
	\frac{(\Gamma_N)_{kaij}}{\sqrt{n_k(1 - n_a)n_in_j}} &= g_y\tilde{\lambda}^k_b(\tau_y)s^{ba}_{ij}(\tau_y)
		+ P(ij)g_y\tilde{\lambda}^k_b(\tau_y)s_i^b(\tau_y)s_j^a(\tau_y) \nonumber \\
	&+ P(ij)\frac{1}{2}g_y\tilde{\lambda}^{kl}_{bd}(\tau_y)s^{bd}_{il}(\tau_y)s_j^a(\tau_y)
	    + P(ij)g_y\tilde{\lambda}^{kl}_{bd}(\tau_y)s^{ad}_{jl}(\tau_y)s_i^b(\tau_y)\nonumber \\
	&- \frac{1}{2}g_y\tilde{\lambda}^{lk}_{db}(\tau_y)s^{db}_{ji}(\tau_y)s_l^a(\tau_y)
		- g_y\tilde{\lambda}^{lk}_{db}(\tau_y)s_j^d(\tau_y)s_l^a(\tau_y)s_j^d(\tau_y) \\
	\frac{(\Gamma_N)_{abij}}{\sqrt{(1 - n_a)(1 - n_b)n_in_j}} &= g_ys_{ij}^{ab}(\tau_y) + \frac{1}{2}P(ij,ab)g_ys_i^a(\tau_y)s_j^b(\tau_y) 
		+ P(ab)g_y\tilde{\lambda}^k_c(\tau_y)s_{ij}^{cb}(\tau_y)s^a_k(\tau_y)\nonumber \\
	&+ P(ij)g_y\tilde{\lambda}^k_c(\tau_y)s^{ab}_{kj}(\tau_y)s_i^c(\tau_y) 
		- P(ij,ab)g_y\tilde{\lambda}^k_c(\tau_y)s^{bc}_{jk}(\tau_y)s_i^a(\tau_y) \nonumber \\ 
	&+ P(ij,ab)g_y\tilde{\lambda}^{k}_c(\tau_y)s_k^a(\tau_y)s_i^c(\tau_y)s_j^b(\tau_y)
	    - \frac{1}{4}g_y\tilde{\lambda}^{kl}_{cd}(\tau_y)s^{ab}_{kl}(\tau_y)s^{cd}_{ij}(\tau_y) \nonumber \\
	&- \frac{1}{2}P(ij,ab)g_y\tilde{\lambda}^{kl}_{cd}(\tau_y)
		s^{ca}_{ki}(\tau_y)s^{db}_{lj}(\tau_y)
	    + \frac{1}{2}P(ab)g_y\tilde{\lambda}^{kl}_{cd}(\tau_y)s^{ac}_{ij}(\tau_y)s^{bd}_{kl}(\tau_y)\nonumber \\
	&+ \frac{1}{2}P(ij)g_y\tilde{\lambda}^{kl}_{cd}(\tau_y)s^{ab}_{ij}(\tau_y)s^{cd}_{jl}(\tau_y)
		- \frac{1}{4}P(ab)g_y\tilde{\lambda}^{kl}_{cd}(\tau_y)s^{cd}_{ij}(\tau_y)s_k^a(\tau_y)s_l^b(\tau_y)
		\nonumber \\
	&- \frac{1}{4}P(ij)g_y\tilde{\lambda}^{kl}_{cd}(\tau_y)s^{ab}_{kl}(\tau_y)s_i^c(\tau_y)s_j^d(\tau_y)
		+ P(ij,ab)g_y\tilde{\lambda}^{kl}_{cd}(\tau_y)s^{bd}_{jl}(\tau_y)s_k^a(\tau_y)s_i^c(\tau_y) \nonumber\\
	&+ \frac{1}{2}P(ij,ab)g_y\tilde{\lambda}^{kl}_{cd}(\tau_y)s^{cd}_{jl}(\tau_y)s_k^b(\tau_y)s_i^a(\tau_y) 
		+ \frac{1}{2}P(ij,ab)g_y\tilde{\lambda}^{kl}_{cd}(\tau_y)s^{bd}_{kl}(\tau_y)s_j^c(\tau_y)s_i^a(\tau_y) \nonumber \\
	&- \frac{1}{4}P(ij,ab)g_y\tilde{\lambda}^{kl}_{cd}(\tau_y)
		s_k^a(\tau_y)s_i^c(\tau_y)s_l^b(\tau_y)sd_j^d(\tau_y) \label{eqn:g2abij}
\end{align}
\endgroup
\end{widetext}
Recall that the indices $i$ and $a$ do not refer to disjoint subspaces and therefore the full unrelaxed density matrix can be written in the MO basis as
\begin{align}
    (\gamma_N)_{pq} &= \sum_{ia}(\gamma_N)_{ia}\delta_{ip}\delta_{aq} +
    \sum_{ba}(\gamma_N)_{ba}\delta_{bp}\delta_{aq} \nonumber\\
    &+ \sum_{ij}(\gamma_N)_{ji}\delta_{jp}\delta_{iq}+ 
    \sum_{ai}(\gamma_N)_{ai}\delta_{ap}\delta_{iq}.
\end{align}
We may compute the average of some operator, $X$, as
\begin{equation}
    \langle X \rangle = \sum_{pq}(\gamma_N)_{qp}X^{(1)}_{pq} + \sum_{pq}p_{qp}X_{pq}
\end{equation}
where $p$ is the mean-field 1-RDM:
\begin{equation}
    p_{qp} = \delta_{qp}n_p.
\end{equation}

The expression for the average of 2-particle properties is analogous. The normal ordered 2-RDM is given by
\begin{equation}
    (\Gamma_N)_{pqrs} = (\Gamma_N)_{ijab}\delta_{ip}\delta_{jq}\delta_{ra}\delta_{sb} + \ldots 
\end{equation}
The full unrelaxed 2-RDM additionally includes the contribution from the reference density:
\begin{align}
    \Gamma_{pqrs} &= (\Gamma_N)_{pqrs} + \frac{1}{2}\left[(\gamma_N)_{pr}p_{qs} - (\gamma_N)_{ps}p_{rq} \right] \nonumber \\
    & + \frac{1}{2}\left[p_{pr}(\gamma_N)_{qs} - p_{ps}(\gamma_N)_{rq} \right] + p_{pr}p_{qs} - p_{ps}p_{qr}.
\end{align}
A 2-electron observable, $Y$, can then be approximated by tracing its operator representation with the unrelaxed 2-RDM:
\begin{equation}
    \langle Y \rangle = \frac{1}{4}\sum_{pqrs}\Gamma_{pqrs}Y_{rspq}
\end{equation}

It is also possible to partially relax the properties by including the response of the orbital energies and the occupation numbers. This involves three contributions:
\begin{equation}
    \frac{\partial \Omega^{(1)}}{\partial n_i}
    \frac{\partial n_i}{\partial \varepsilon_i}
    \frac{\partial \varepsilon_i}{\partial \alpha}, \quad
    \frac{\partial \mathcal{L}}{\partial n_i}
    \frac{\partial n_i}{\partial \varepsilon_i}
    \frac{\partial \varepsilon_i}{\partial \alpha}, \quad
    \frac{\partial \mathcal{L}}{\partial \varepsilon_i }
    \frac{\partial \varepsilon_i}{\partial \alpha}.
\end{equation}
Since the derivative of the orbital energies is just equal to 
\begin{equation}
    \frac{\partial \varepsilon_i}{\partial \alpha} = \matrixel{i}{X}{i}
\end{equation}
for a 1-electron operator, the contribution due to the relaxation of the orbital energies and occupation numbers can be computed as
\begin{equation}
    \sum_q d_{q}X_{q}^{(0)}
\end{equation}
where 
\begin{align}
    d_q &= \frac{\partial \Omega^{(1)}}{\partial n_q}
    \frac{\partial n_q}{\partial \varepsilon_q} + 
    \frac{\partial \mathcal{L}}{\partial n_q}
    \frac{\partial n_q}{\partial \varepsilon_q} + \frac{\partial \mathcal{L}}{\partial \varepsilon_q}.
\end{align}
The first term is most easily computed from the derivative of the first order correction to $\Omega$:
\begin{align}
    \frac{\partial \Omega^{(1)}}{\partial n_q} &= (h_{qq} - \varepsilon_q) + 
    \sum_j\left[\matrixel{qj}{V}{qj} - \matrixel{qj}{V}{jq}\right]n_j \nonumber\\
    &= \matrixel{q}{f}{q} - \varepsilon_q.
\end{align}
This contribution will be zero for a thermal Hartree-Fock reference.

The second term is most efficiently computed by first forming derivative integrals,
\begin{align}
    f^{(q)}_{ab} &= \frac{\partial n_q}{\partial \varepsilon_q}\frac{\partial}{\partial n_q}f_{ab} \\
    \bra{ab}\ket{cd}^{(q)} &= \frac{\partial n_q}{\partial \varepsilon_q}\frac{\partial}{\partial n_q}\bra{ab}\ket{cd},
\end{align}
and then contracting them with the unrelaxed, normal-ordered, 1- and 2-RDMs:
\begin{equation}
    (\gamma_N)_{ba}f^{(q)}_{ab} + (\Gamma_N)_{cdab}\bra{ab}\ket{cd}^{(q)}
    + \ldots
\end{equation}
Note that while the derivatives of the Fock matrix are dense 3-index quantities because the Fock matrix involves sums over occupation numbers, the derivatives of the two-electron interaction are still only 4-index quantities because
\begin{align}
    \bra{ab}\ket{cd}^{(q)} &= \bra{ab}\ket{cd}^{(a)}\delta_{qa}
    + \bra{ab}\ket{cd}^{(b)}\delta_{qb}\nonumber\\
    &+ \bra{ab}\ket{cd}^{(c)}\delta_{qc}
    + \bra{ab}\ket{cd}^{(d)}\delta_{qd}
\end{align}
The remaining contribution is computed by taking the derivative of orbital energies which appear directly in the Lagrangian in the exponential factor. If the integral form of the equations are used, then this term can be computed directly:
\begin{align}
    \frac{\partial \mathcal{L}}{\partial \varepsilon_q } &= -
	\frac{1}{\beta}\sum_y g_y \lambda^{\mu}(\tau_y)\nonumber \\
	&\qquad \times \sum_x (\tau_y - \tau_x)G_x^y e^{\Delta_{\mu}(\tau_x - \tau_y)} \mathrm{S}_{\mu}(\tau_x)\frac{\partial \Delta_{\mu}}{\partial \varepsilon_q}
\end{align}
The label $\mu$ runs over all singles and doubles, and the derivatives of the energies differences are sparse in that
\begin{equation}
    \frac{\partial\Delta_i^a}{\partial \varepsilon_q} = \delta_{aq} - \delta_{iq}. 
\end{equation}
In the case that the differential form of the equations is used, this term must be calculated as
\begin{equation}
    \frac{1}{\beta}\int_0^{\beta}d\tau \text{S}_{\mu}[\mathbf{s}(\tau)]\frac{\partial\tilde{\lambda}^{\mu}(\tau)}{\partial\varepsilon_p}
\end{equation}
where the derivative appearing under the integral can be propagated along with $\tilde{\lambda}$
\begin{equation}
    \frac{d}{d\tau}\frac{\partial\tilde{\lambda}^{\mu}(\tau)}{\partial\varepsilon_p} = \Delta_{\mu}\frac{\partial\tilde{\lambda}^{\mu}(\tau)}{\partial\varepsilon_p} + \tilde{\lambda}(\tau).
\end{equation}
Usually, $X^{(0)}$ contains all the diagonal elements and $X^{(1)}$ contains all the off-diagonal elements. In this case, we can construct one partially-relaxed FT-CCSD density matrix as
\begin{equation}
    p^{cc}_{qp} = (\gamma_N)_{qp}(1 - \delta_{qp}) + \delta_{qp}(d_q + n_q).
\end{equation}
This is the FT-CCSD ``density matrix" which incorporates the relaxation of the orbital energies and occupation numbers. 

We will not explicitly discuss the procedure for including orbital response (term 3), as we do not consider these terms in this work. However, the computation of the FT-CC Z-vector parallels closely the ground-state case which is discussed in Ref.~\onlinecite{Salter1989}.

For the derivatives with respect to $\beta$, there are 3 additional terms that we must consider. Term 4 is just
\begin{equation}\label{eqn:diff4}
    (4) = -\frac{1}{\beta}\Omega_{cc}.
\end{equation}
Term 5 can be computed specifically for a particular discretization by evaluating the Lagrangian with 
\begin{equation}
    \frac{\partial G}{\partial \beta} \quad \text{and} \quad 
    \frac{\partial g}{\partial \beta}
\end{equation}
respectively, or these terms can be computed as the derivative of the integration limits in the Lagrangian. This amounts to evaluating the integrand of $\mathcal{L}$ at $\tau = \beta$, and two methods will agree in the limit of a dense grid. Terms of type 6 are simple to write down due to the fact that the positions of the grid points depend linearly on $\beta$,
\begin{equation}
    \frac{\partial \tau_y}{\partial \beta} = \frac{\tau_y}{\beta},
\end{equation}
therefore
\begin{align}
    (6) &= -\frac{1}{\beta^2}\sum_y g_y \lambda^{\mu}(\tau_y)\nonumber \\
    & \qquad \times \sum_x \Delta_{\mu} (\tau_y - \tau_x)G_x^y e^{\Delta_{\mu}(\tau_x - \tau_y)} \mathrm{S}_{\mu}(\tau_x).
    \label{eqn:diff6}
\end{align}
This final term will vanish in the limit of a dense grid and can therefore be ignored without affecting the properties in the limit as $n_g \rightarrow \infty$.

\section{Entropy of the 1D Hubbard model}\label{sec:HubS}
\begin{figure}[!ht]
    \centering
    \includegraphics{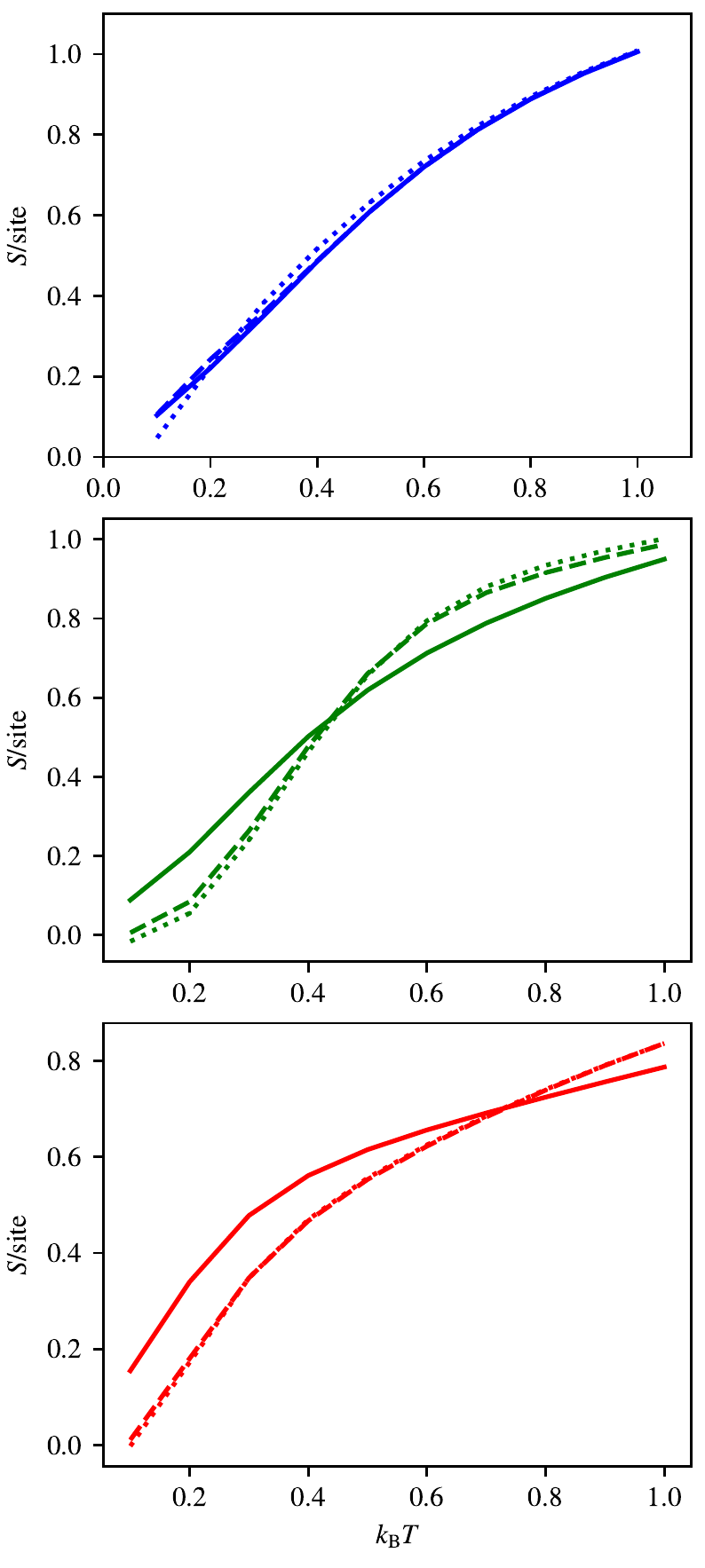}
    \caption{More detailed view of the entropy per site of the 1D Hubbard model at half filling for $U = 2$ (blue, top), $U = 4$ (green, middle), and$U = 8$ (red, bottom). This is the same as row 3 (1st column) of Figure~\ref{fig:hubbard}.}
    \label{fig:HubS}
\end{figure}

In Figure~\ref{fig:HubS}, the entropy of the 1D Hubbard model is plotted in more detail.
FT-CCSD consistently underestimates the entropy at low temperatures, and this effect is more pronounced at for larger $U$.

\section{Basis set error in the warm, dense UEG}\label{sec:UEG_basis}
The exchange-correlation energy is defined for a fixed number of electrons. In the grand canonical ensemble, we compute $E_{xc}$ for $N$ electrons in $M$ plane wave orbitals as
\begin{align}
    E_{xc}(N,M) &= E_{CC}(\mu_{CC}(M), M) + E_0(\mu_{CC}(M), M) \nonumber \\
    &- E_0(\mu_0(M), M)
\end{align}
where $E_{CC}$ is the FT-CCSD exchange-correlation energy and $\mu_{CC}$ and $\mu_0$ are chosen separately so that the coupled cluster and reference systems each have $N$ electrons. The simplest basis extrapolation technique, which we will refer to as ``E1," to extrapolate this quantity assuming that the basis dependence behaves asymptotically like $1/M$. However, at higher temperatures, there will be significant finite-basis error in the computation of $E_0$ and we could also compute 
\begin{align}
    E_{xc}'(N,M) &= E_{CC}(\mu_{CC}(M), M) + E_0(\mu_{CC}(M), \infty) \nonumber \\
    &- E_0(\mu_0(M), \infty).
\end{align}
The extrapolation of this quantity based on an asymptotic $1/M$ dependence will be referred to as ``E2." Other types of extrapolations are possible, but these two are sufficient for our purposes.

\begin{figure}[!ht]
    \centering
    \includegraphics{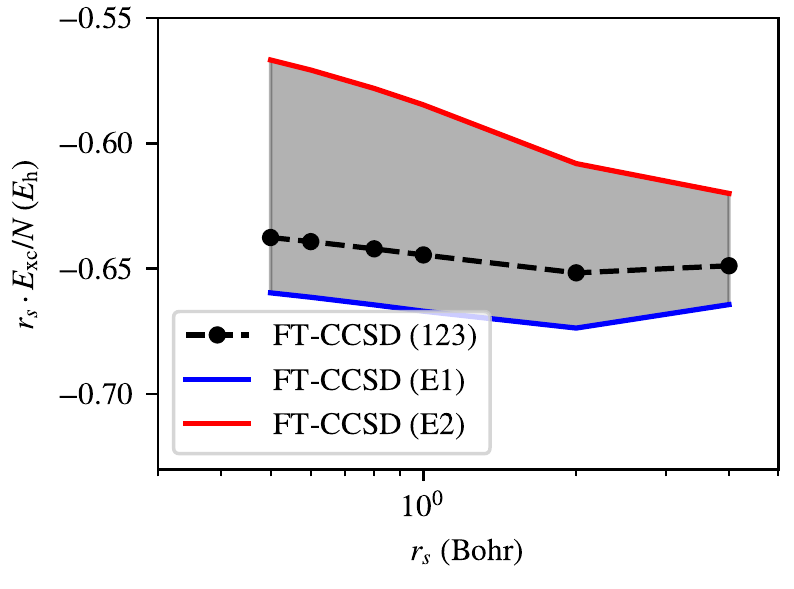}
    \caption{The FT-CCSD exchange correlation energy of the $N = 33$ polarized UEG at a reduced temperature of $\theta = 0.5$. The exchange-correlation energy is scaled by $r_s$ to make the scale of the plot more uniform. The solid lines is the extrapolated values based on the E1 (blue) and E2 (red) methods. The shaded region provides a rough estimate of the uncertainty in these extrapolations.}
    \label{fig:PUEG_050_basis}
\end{figure}

\begin{figure}[!ht]
    \centering
    \includegraphics{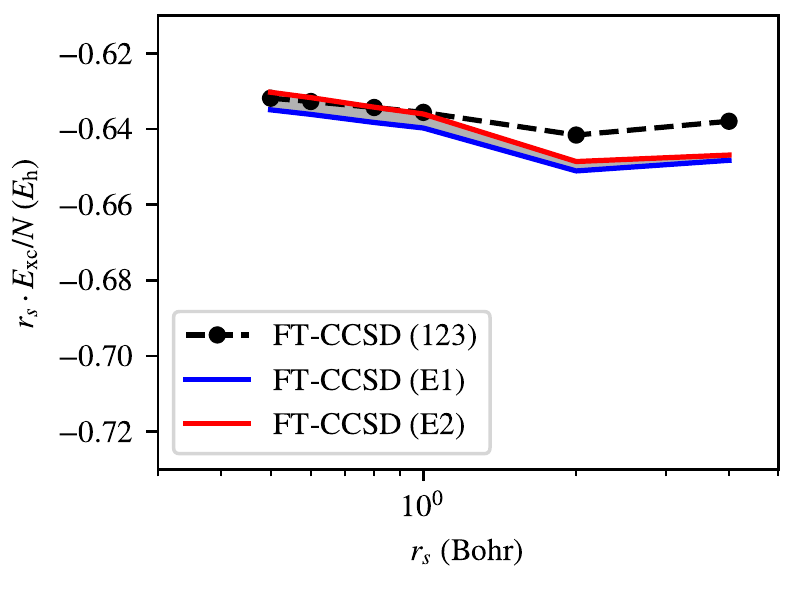}
    \caption{The FT-CCSD exchange correlation energy of the $N = 33$ polarized UEG in two different plane-wave basis sets at a reduced temperature of $\theta = 0.25$. The exchange-correlation energy is scaled by $r_s$ to make the scale of the plot more uniform. The solid line is the extrapolated value.}
    \label{fig:PUEG_025_basis}
\end{figure}

In Figures~\ref{fig:PUEG_050_basis} and~\ref{fig:PUEG_025_basis} we plot the FT-CCSD exchange correlation energy of the polarized UEG in a basis set of 123 plane waves. Additionally, we have extrapolated to the complete basis set limit with the E1 and E2 methods using basis set sizes of 93 and 123 plane waves. The difference in the two extrapolations, which should provide the same answer asymptotically, allows us to estimate the uncertainty in the basis set extrapolation. For $\theta = 0.5$ the uncertainty is quite large, and there is no reason to think that either extrapolation is more reliable than the $M = 123$ results. On the other hand, for $\theta = 0.25$ both E1 and E2 methods provide similar results which suggests that either may provide a better estimate than the $M = 123$ results.

\begin{figure}[!ht]
    \centering
    \includegraphics{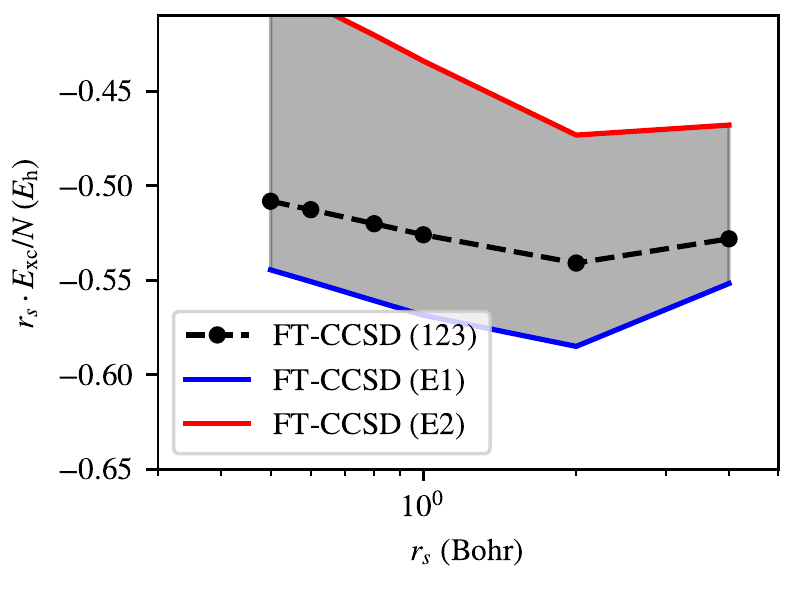}
    \caption{The FT-CCSD exchange correlation energy of the $N = 66$ unpolarized UEG in a basis set of 123 plane-wave basis orbitals at a reduced temperature of $\theta = 0.5$. The exchange-correlation energy is scaled by $r_s$ to make the scale of the plot more uniform. E1 and E2 extrapolations are plotted in blue and red respectively.}
    \label{fig:UEG_050_basis}
\end{figure}

\begin{figure}[!ht]
    \centering
    \includegraphics{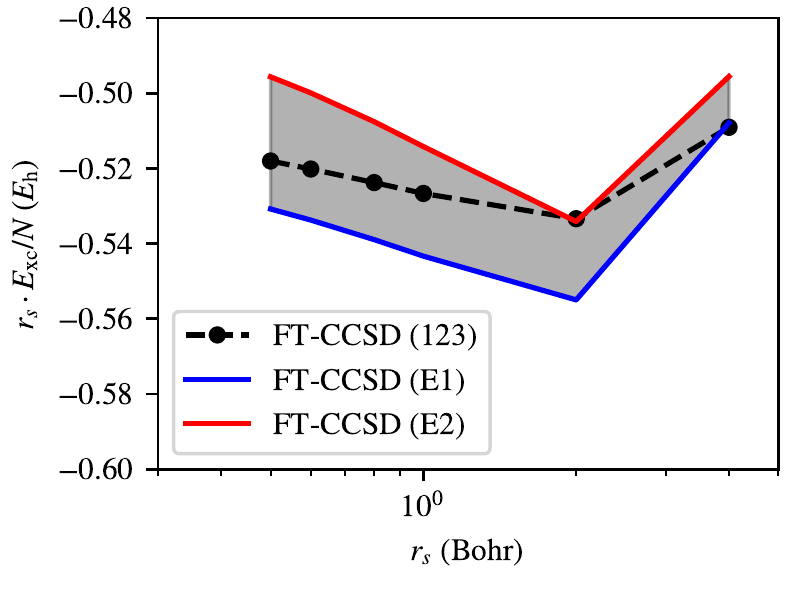}
    \caption{The FT-CCSD exchange correlation energy of the $N = 66$ polarized UEG in a basis set of 123 plane-wave basis orbitals at a reduced temperature of $\theta = 0.25$. The exchange-correlation energy is scaled by $r_s$ to make the scale of the plot more uniform. E1 and E2 extrapolations are plotted in blue and red respectively.}
    \label{fig:UEG_025_basis}
\end{figure}

In Figures~\ref{fig:UEG_050_basis} -~\ref{fig:UEG_025_basis} we plot the FT-CCSD exchange correlation energy of the unpolarized UEG in a basis set of 123 plane waves along with the results of E1 and E2 extrapolations.
As with the polarized UEG, there is a larger difference between the E1 and E2 methods at higher temperature. This makes sense because at higher temperature states with larger kinetic energy will be thermally populated, and a larger plane-wave basis will be necessary. Unlike for the polarized UEG, the basis set extrapolation is probably not reliable at either temperature. Calculations in larger basis sets should allow for basis set extrapolation with greater confidence.

\section*{References}
\bibliography{ftccsdII}
\end{document}